\documentclass[useAMS,usenatbib]{mn2e}
\usepackage{aas_macros}
\usepackage{graphicx}
\usepackage{amsmath,amssymb}
\usepackage[T1]{fontenc}
\usepackage{aecompl}
\title[Supernova Feedback in an Inhomogeneous Interstellar Medium]{Supernova Feedback in an Inhomogeneous Interstellar Medium}
\author[D. Martizzi et al.]{\parbox[t]{\textwidth}{Davide Martizzi$^{1}$\thanks{E-mail: davide.martizzi@berkeley.edu}, 
Claude-Andr\'{e} Faucher-Gigu\`{e}re$^{2}$, Eliot Quataert$^{1}$}\vspace*{6pt}\\
$^{1}$Department of Astronomy and Theoretical Astrophysics Center, University of California, Berkeley, CA 94720-3411, USA.\\
$^{2}$Department of Physics \& Astronomy and Center for Interdisciplinary Exploration and Research in Astrophysics (CIERA),\\ Northwestern University, Northwestern University, Evanston, IL 60208-3112, USA.\\
}

\begin{document}

\maketitle

\label{firstpage}

\begin{abstract}

Supernova (SN) feedback is one of the key processes shaping the interstellar medium (ISM) of galaxies.  SNe contribute to (and in some cases may dominate) driving turbulence in the ISM and accelerating galactic winds.  
Modern cosmological simulations have sufficient resolution to capture the main structures in the ISM of galaxies, but are typically still not capable of explicitly resolving all of the small-scale stellar feedback processes, 
including the expansion of supernova remnants (SNRs).  We perform a series of controlled three-dimensional hydrodynamic (adaptive mesh refinement) simulations of single SNRs expanding in an inhomogeneous density field 
with statistics motivated by those of the turbulent ISM.  We use these to quantify the momentum and thermal energy injection from SNe as a function of spatial scale and the density, metallicity, and structure of the 
ambient medium. We develop a series of analytic formulae that we fit to the simulations. These formulae can be used as a basis for a more predictive sub-resolution model for SN feedback for galaxy formation simulations. 
We then use simulations of multiple, stochastically driven SNe that resolve 
the key phases of SNRs to test the sub-resolution model, and show that it accurately captures the turbulent kinetic energy and thermal energy in the ISM. 
By contrast, proposed SN feedback models in the literature based on `delayed cooling' significantly overpredict the late-time thermal energy and momentum 
in SNRs. 
\end{abstract}


\section{Introduction}
Supernovae are some of the most energetic events within galaxies. Each SN injects in the ISM a kinetic energy of $\sim 10^{51}$ erg in the form a few solar masses of stellar ejecta moving initially at $\sim 10^4$ km s$^{-1}$.  
SNe contribute significantly to, and in some cases may dominate, driving interstellar turbulence \citep[e.g.,][]{2006ApJ...638..797D, 2006ApJ...653.1266J, 2009ApJ...704..137J, 2011ApJ...731...41O, 2013MNRAS.433.1970F} and 
accelerating galactic winds from star-forming galaxies \citep[e.g.,][]{2005ARA&A..43..769V, 2009ApJ...697.2030S, 2012MNRAS.421.3522H}.  SNe also strongly influence the dynamics and phase structure of the ISM by inflating 
bubbles of hot gas \citep[e.g.,][]{1977ApJ...218..148M} and by accelerating relativistic cosmic rays \citep[e.g.,][]{1978ApJ...221L..29B}.

Galaxy formation models that do not include strong stellar feedback lead to galaxies that convert their gas into stars too rapidly by a factor $\sim100$ \citep[e.g.,][]{2011MNRAS.417..950H, 2013ApJ...770...25A} relative 
to observations \citep[][]{1998ApJ...498..541K, 2010MNRAS.407.2091G}.  They also form too many stars overall by a factor 
$\sim5$ to $>10^{3}$ \citep[e.g.,][]{1991ApJ...379...52W, 2009MNRAS.396.2332K, 2010ApJ...710..903M, 2010ApJ...717..379B, 2011MNRAS.417.2982F} and fail to explain the observed distribution of heavy elements in 
the intergalactic medium \citep[e.g.,][]{2001ApJ...561..521A, 2006MNRAS.373.1265O, 2010MNRAS.409..132W}.  Stellar feedback in general, and supernova feedback in particular, is thus an essential ingredient in galaxy 
formation (e.g., \citealt{Dekel1986}).  Approximations to the impact of SNe on the ISM are included in most modern simulations 
\citep[e.g.,][]{2011ApJ...742...76G, 2013ApJ...770...25A, 2012MNRAS.421.3488H, 2014MNRAS.445..581H, 2014MNRAS.437.1750M}.

However, many implementations of stellar feedback utilize \emph{ad hoc} approximations intended to limit radiative losses and ensure that the feedback is sufficiently strong to reproduce the properties of observed 
galaxies.  These approximations include temporarily suppressing hydrodynamical interactions \citep[e.g.,][]{2003MNRAS.339..312S, 2006MNRAS.373.1265O, 2013MNRAS.436.3031V} or gas cooling 
\citep[e.g.,][]{2006MNRAS.373.1074S, 2007MNRAS.374.1479G, 2010MNRAS.407.1581S, 2011ApJ...742...76G} as kinetic or thermal energy is injected to model stellar feedback. In addition to being inaccurate in detail (e.g., in the phase structures that they predict for the ISM and galactic winds), these approximations often have tunable parameters, which limits their predictive power.

In this paper, we perform a series of high-resolution simulations of isolated SNRs aimed at quantifying the momentum and thermal energy injected in the ISM by SNe.  A vast literature is already available on SNR evolution, 
but most previous calculations have assumed that the ambient medium is uniform \citep[][]{1974ApJ...188..501C, 1988ApJ...334..252C, 1991ApJ...383..621D, 1998ApJ...500...95T}.\footnote{Several studies of stellar feedback 
\citep[e.g.,][]{2005ApJ...630..167T, 2013MNRAS.432..455D} have assumed that the momentum injected in the ISM by SNe scales with ambient medium density as $P_{\rm fin}\propto n_{\rm H}^{-0.25}$ following the fitting formulae 
of \cite{1998ApJ...500...95T} for the swept up mass and outer shock velocity.  \cite{2013MNRAS.433.1970F} noted that this scaling is inconsistent with the weaker scaling $P_{\rm fin} \propto n_{\rm H}^{-1/7}$ indicated by 
other analytic models and numerical simulations of SNR evolution \cite[e.g.,][]{1988ApJ...334..252C, 1991ApJ...383..621D, 1998ApJ...500..342B}.  They suggested that the problem was \cite{1998ApJ...500...95T}'s fit for the 
shock velocity.  In this paper, we confirm the weaker scaling with ambient density for a homogeneous medium, but show that the scaling is intermediate for the inhomogeneous case, $P_{\rm fin} \propto n_{\rm H}^{-0.19}$ 
(equation (\ref{Pfin inhomo})).  Using the correct scaling is important when evaluating the relative importance of, e.g., SNe and radiation pressure in driving ISM turbulence in dense galaxies.  }  We expand on these 
previous studies by systematically studying the case of an inhomogeneous ambient medium with statistics motivated by the supersonically turbulent gas observed in the ISM.  Specifically, we use as initial conditions a 
density field with a lognormal probability density function (PDF) and a Burgers spatial power spectrum and determine how the evolution of the SNR depends on the density structure and metallicity of the ambient medium.

Using these simulation results, we derive simple fitting formulae for the key quantities as a function of radius during the expansion of a SNR.   We then use our fitting formulae to construct a sub-resolution model 
for SN feedback for use in galaxy-scale simulations that do not necessarily resolve the key evolutionary phases of SNRs (e.g., the Sedov-Taylor phase).   Such a model is important for a number of reasons.  
For example, during the Sedov-Taylor phase, energy $\propto M v^{2}$ is conserved while the momentum $\propto M v$ of the remnant increases by a factor $\sim5-30$ due to work done by hot shocked gas.
Thus, if the Sedov-Taylor phase of SNRs is not accurately captured in simulations the momentum injected in turbulence may be underestimated by at least one order of magnitude.\footnote{If SNe are modeled via thermal 
energy injection only and the energy is completely radiated away due to inadequate resolution, the net momentum injection is zero.} 

In the FIRE (Feedback In Realistic Environments) cosmological simulation project\footnote{See the FIRE project website at: http://fire.northwestern.edu.} \citep[][]{2014MNRAS.445..581H}, a subset of the authors have 
implemented a model in which SN explosions are sources of thermal energy when the cooling radius is well resolved (so that the SNR evolution is self-consistently captured by the simulation) but primarily sources of 
radial momentum when the cooling radius is not resolved.  In that work, the cooling radius and asymptotic radial momentum and thermal energy were determined using spherically-symmetric models of SNRs in a homogeneous 
medium.  A primary goal of the present paper is to improve the accuracy of this type of model by quantifying the evolution of SNRs in a realistic inhomogeneous medium.  In addition, we explicitly demonstrate the validity 
of our sub-grid model by comparing its predictions for multiple stochastically generated (in space and time) SNe in a periodic box with analogous simulations that explicitly resolve the key evolutionary phases of SNRs. 
{Our implementation of the sub-grid model in the multiple supernova simulations is intentionally simplified: effects related to the clustering of supernovae in space and time are 
neglected. Clustering can play an important role in stratified media (e.g. galactic disks). For example, it can promote the formation of hot superbubbles, potentially reducing the amount energy that is converted into 
turbulence at the midplane in favor of driving a galactic wind or a galactic fountain \citep{1977ARA&A..15..175C, 1977ApJ...218..377W, 1989ApJ...337..141M, 1999ApJ...513..142M, 2014MNRAS.442.3013K, 2014MNRAS.443.3463S}. Clustering effects are beyond the scope of the present paper, but we plan to study them in a future paper that will also include 
the effects of vertical stratification.}

This paper is structured as follows. \S \ref{sec:num_methods} describes the numerical methods and the simulations we run. 
\S \ref{sec:isolated_SNRs} describes our results on the evolution of isolated SNRs. 
In \S \ref{sec:SNR_evolution}, we summarize the overall evolution of  SNRs in an inhomogeneous medium and compare the results to the more widely studied problem of SNe in a homogeneous medium. 
\S \ref{sec:mom_egy_isolated} focuses on the momentum and thermal energy evolution of individual SNRs. 
In \S \ref{sec:fits} we derive analytic approximations to our numerical results and in \S \ref{sec:subgrid} we discuss how these analytic approximations can be implemented as a sub-grid model  in lower resolution simulations.
In \S \ref{sec:testing_subgrid}, we test the sub-resolution model using simulations of multiple SNRs stochastically driven and interacting self-consistently in a periodic box. 
\S \ref{sec:summary} summarizes and discusses our results.

\section{Numerical simulations}
\label{sec:num_methods}
All of the simulations have used the {\scshape ramses} code \citep{Teyssier:2002p451}, an adaptive mesh refinement 
(AMR) code based on a second-order unsplit Godunov solver \citep{Toro:1994p1151}. We evolve an ideal hydrodynamic fluid in 3D without self-gravity. 
To ensure numerical stability, we adopt the Local Lax-Friedrichs (LLF) Riemann solver. 
We model radiative cooling using the metallicity-dependent cooling function of \cite{1993ApJS...88..253S}. This cooling function does not include fine structure metal line or molecular line cooling and so is 
effectively truncated below $T\approx10^4$ K.

The evolution of a SNR depends on the structure of the ambient ISM. 
For this reason, we run three types of simulations: 
\begin{enumerate}
\item isolated SNRs in a homogeneous ISM; 
\item isolated SNRs in an inhomogeneous ISM; 
\item multiple SNRs interacting self-consistently in a periodic box.
\end{enumerate}
The simulations in a homogeneous medium are used to compare our results with previously published solutions \citep{1974ApJ...188..501C, 1988ApJ...334..252C, 1991ApJ...383..621D, 1998ApJ...500...95T}.  
Our simulations of isolated SNRs in an inhomogeneous medium allow us to determine how the evolution is modified in a more realistic ISM.  We use our simulations of isolated SNRs to develop a sub-resolution model for 
galaxy-scale simulations with insufficient resolution to explicitly capture the cooling radius of SNRs.  Finally, we test this sub-grid model using the simulations of multiple SNRs.

\subsection{Isolated SNRs in a homogeneous ISM }
For this set of simulations, we consider cubic boxes filled with a homogeneous ISM initially in pressure equilibrium.  SNe are modeled by injecting mass (SN ejecta with $M_{\rm ej}=3$ M$_{\odot}$), thermal energy 
($E_{\rm th} = 6.9 \times 10^{49}$ erg), and radial kinetic energy ($E_{\rm kin} = 9.31 \times 10^{50}$ erg) in a spherical region of radius $7 \Delta x$ (where $\Delta x$ is the cell size) at the centre of each box.  
The total energy injected is $E_{\rm tot}=E_{\rm th}+E_{\rm kin}=10^{51}$ erg. These initial condition are the same as those of \cite{1998ApJ...500...95T}.  They do not correspond exactly to the partitioning of thermal 
and kinetic energy in the Sedov-Taylor phase, which has somewhat more thermal energy.  However, the Sedov-Taylor solution is an attractor and the SNRs in our homogeneous ISM simulations quickly adjust to match the 
Sedov-Taylor solution (see Figure \ref{hom_vs_inh} below).
 
The evolution of an isolated SNR can be characterized by several key phases:
\begin{enumerate}
\item the free expansion phase, during which the mass of the SN ejecta is larger than the mass of the swept up gas; 
\item the Sedov-Taylor phase, during which radiative losses are negligible and the total energy in the remnant is conserved \citep[in this phase, $E_{\rm th}=0.717 E_{\rm tot}$;][]{1988ApJ...334..252C};
\item the pressure-driven snowplow phase, during which radiative losses begin to influence the  remnant evolution and the expansion of the blast wave is determined by the pressure in the shocked gas; 
\item the momentum-conserving snowplow phase; this begins when sufficient energy has been radiated away that the momentum of the swept up gas reaches its asymptotic value. 
\end{enumerate}
The radius at which radiative losses become important is called the cooling radius (this is quantified in \S \ref{sec:mom_egy_isolated}). {In a very hot and low density medium ($n_{\rm H}<10^{-2}$ cm$^{-3}$) 
the SNRs likely never reach the snowplow phase, as their shells do not cool prior to slowing to the local (fast) sound speed \citep{1988ApJ...324..776M}. In such cases the SNR is expected to dissolve in the ISM before 
significant radiative losses can be achieved.}

We use previous solutions of SNRs in a homogeneous medium \citep{1988ApJ...334..252C, 1998ApJ...500...95T} to approximate how the cooling radius scales with properties of the ambient medium:  
$R_{\rm c}\approx14.0 \hbox{\rm~pc}\left( n_{\rm H}/1 \hbox{ cm}^{-3}\right) ^{-3/7} \left( E_{\rm tot}/ 10^{51} \hbox{ erg}\right) ^{2/7} \left( Z/Z_{\odot}\right) ^{-1/7}$.  We then set the box size and resolution in our 
simulations so that the cooling radius is always resolved by at least 50 grid cells and so that the box is always large enough to capture the evolution in the momentum-conserving phase (2-3 cooling radii are sufficient to 
capture the momentum-conserving phase). 

We use AMR to speed up our calculations. 
We adopt a refinement scheme based on pressure gradients which refines around shocks. 
The maximum resolution we achieve in the fiducial runs is equivalent to that obtained with a Cartesian grid with $512^3$ cells (maximum refinement level 
$\ell_{\rm max} = 9$). We consider four different ambient medium gas densities: $n_{\rm H} = 0.01,~1,~100$ and $10^4$ cm$^{-3}$. We adopt a fiducial gas metallicity $Z = Z_{\odot}$ in 
all the four cases. For the case $n_{\rm H} = 100$ cm$^{-3}$, we run two additional simulations with different metallicities $Z=0.1 Z_{\odot}$ and $Z=0.01 Z_{\odot}$. 
The parameters of our isolated SNR simulations are summarized in Table~\ref{tab:parameters}. 
The runs with homogeneous ambient medium are prefixed by `h\_.'

\subsection{Isolated SNRs in an inhomogeneous ISM }\label{sec:iso_sims_descr}
The real ISM is highly inhomogeneous \citep{1977ApJ...218..148M, 2004ARA&A..42..211E, 2009ApJ...704..137J,  
2012MNRAS.421.3488H}. 
For our simulations of isolated SNRs in an inhomogeneous medium, we do not attempt to self-consistently generate a realistic multiphase ISM. 
Rather, we use parametric initial conditions with statistics motivated by those of supersonic turbulence that enable us to systematically study how the SNR evolution depends on the properties of the ambient medium. 

We use the same Riemann solver, effective resolution (i.e. AMR with the same levels of refinement), and box sizes as for the simulations with a homogeneous ISM, and launch SNRs by injecting the same initial energy and momentum. 
The inhomogeneous medium is also initialized in pressure equilibrium; we ensure that the temperature in each cell is below $10^4$ K so that the gas does not initially cool. The density PDF is a lognormal 
\begin{equation}\label{eq:lognormal}
 f(y) dy = \frac{1}{\sqrt{2 \pi \sigma^2}}\exp\left[\frac{-(y+\mu)^2}{2 \sigma^2}\right]dy,
\end{equation}
where $y=\ln{n/\bar{n}}$ and $\bar{n}$ is the mean density. The parameters $\mu$ and $\sigma$ are related to the properties of the turbulence in the ISM. We adopt the parameterization of 
\cite{2009RMxAC..36..243L} which is appropriate for hydrodynamic turbulence:
\begin{equation}\label{PDF:2}
 \mu=-0.36\log_{\rm 10}(1 + 0.5\mathcal{M}^2) + 0.10
\end{equation}
and
\begin{equation}\label{PDF:3}
 \sigma^2=2|\mu|,
\end{equation}
where $\mathcal{M}$ is the Mach number of the turbulence. 
Spatial correlations in the initial conditions are parameterized by a 3D power spectrum,
\begin{equation}\label{powspec}
 P(k)\propto k^{-\beta},
\end{equation}
between $k_{\rm min}$ and $k_{\rm max}$, and zero otherwise. 
In what follows, $\lambda =1/k$ is the spatial scale of a perturbation. 
For our fiducial simulations, we adopt a Burgers power spectrum with $\beta=4$. {In reality, the density distribution in the ISM is not only set by the properties of isothermal 
turbulence but also by physical processes such as the thermal instability. In certain circumstances, the true density PDF is therefore broader than the turbulence model described 
above would suggest \citep{2004Ap&SS.289..479D}. We do not include these additional physical processes explicitly in this work, though their effects can be approximated by considering 
higher Mach numbers (which also produce broader density PDFs). }

We use an iterative method to generate a lognormal random field with the power spectrum in equation~(\ref{powspec}). 
Specifically, we iterate to find the the power spectrum for the corresponding Gaussian random field, which we then exponentiate to obtain the lognormal field. The iterative method is based on \cite{2002LewisAustin} 
(see also \citealt{2007ApJS..173...37S}). The normalization of the power spectrum is set by fixing the value of $\sigma$, i.e. by the Mach number of the turbulent motions. 

The maximum scale $\lambda_{\rm max}$ with finite power is fiducially set to be $L_{\rm box}/15$. We keep $\lambda_{\rm min}=L_{\rm box}/128$ fixed but explore different values of 
$\lambda_{\rm max}$ to test how the evolution of a SNR is influenced by being in an unusually underdense/overdense region. 
If $\lambda_{\rm max}$ is small compared to the cooling radius, the blast wave 
in general sweeps up many outer-scale density fluctuations before reaching the cooling radius.  However if $\lambda_{\rm max}$ is large compared to the cooling radius, the blast wave evolution will 
depend on the location of the SN because it will depend on whether the SN explodes in an underdense or overdense region.  This is, of course, physically realistic.  In our calculations, however, 
we explicitly vary the mean ambient density as a separate parameter and choose  $\lambda_{\rm max} \lesssim R_c$ to ensure that the results of the SNR evolution are statistically roughly independent of where the 
SN is placed.  The magnitude of the density fluctuations in our simulations are thus best interpreted as the density fluctuations averaged over the cooling radius (not the density fluctuations on the outer-scale 
in the turbulent ISM).

We run a set of simulations in which we vary the mean density, metallicity, Mach number and the maximum scale of the density fluctuations in the ambient medium. 
We consider four values for the mean density: 
$\bar{n}_{\rm H} = 0.01,~1,~100$ and $10^4$ cm$^{-3}$. 
For $\bar{n}_{\rm H} = 100$ cm$^{-3}$, we explored several parameter variations: Mach number $\mathcal{M}=1, 10, 30$; metallicity $Z=0.01, 0.1, 1.0$ $Z_{\odot}$; 
and $\lambda_{\rm max} = L_{\rm box}/15, L_{\rm box}/5, L_{\rm box}/2.5$. The fiducial metallicity is $Z=Z_{\odot}$, the fiducial Mach number is $\mathcal{M}=30$, and the fiducial maximum spatial scale of density 
fluctuations is $\lambda_{\rm max} = L_{\rm box}/15$. 

We set the initial velocity field of the ambient medium to zero. Thus, the ambient medium is not actually turbulent but it does have the density fluctuations of a turbulent medium. This is a good approximation because 
for realistic Mach numbers the turbulent velocities are typically much smaller than the speed of the SNR, except at the very end of the SNR evolution when the SNR merges with the ISM. 

The simulations with inhomogeneous ISM are identified by the prefix `i$\_$'.  Their parameters are summarized in Table~\ref{tab:parameters}.  We discuss the results of these simulation in \S \ref{sec:isolated_SNRs}. 
Resolution tests confirm that our solutions are numerically converged (see Appendix~\ref{appendix:A}).

\begin{table*}
\begin{center}
{\bfseries Parameters of isolated SNR simulations}
\end{center}
\begin{center}
\begin{tabular}{lccccccc}
\hline
 Name & Medium Type & $\bar{n}_{\rm H}~[{\rm cm}^{-3}]$ & $Z~[Z_{\odot}]$ & $\mathcal{M}$ & $L_{\rm box}~[{\rm pc}]$ & $\Delta x~[{\rm pc}]$ & $k_{\rm min}=1/\lambda_{\rm max}~[{\rm pc^{-1}}]$ \\
\hline
h\_n1e-2\_z1 & Homogeneous & 0.01 & 1 & n.a. & 2000 & 3.9 & n.a. \\
h\_n1\_z1 & Homogeneous & 1 & 1 & n.a. & 128 & 0.25 & n.a. \\
h\_n100\_z0.01 & Homogeneous & 100 & 0.01 & n.a. & 18.5 & 0.036 & n.a. \\
h\_n100\_z0.1 & Homogeneous & 100 & 0.1 & n.a. & 18.5 & 0.036 & n.a. \\ 
h\_n100\_z1 & Homogeneous & 100 & 1 & n.a. & 18.5 & 0.036 & n.a. \\
h\_n1e4\_z1 & Homogeneous & 10$^{4}$ & 1 & n.a. & 2.67 & 0.0052 & n.a. \\
i\_n1e-2\_z1 & Inhomogeneous & 0.01 & 1 & 30 & 2000 & 3.9 & $15/L_{\rm box}=0.0075$ \\
i\_n1\_z1\_M30 & Inhomogeneous & 1 & 1 & 30 & 192 & 0.375 & $15/L_{\rm box} = 0.078$ \\
i\_n100\_z0.01\_M30 & Inhomogeneous & 100 & 0.01 & 30 & 27.75 & 0.054 & $15/L_{\rm box} = 0.54$ \\
i\_n100\_z0.1\_M30 & Inhomogeneous & 100 & 0.1 & 30 & 27.75 & 0.054 & $15/L_{\rm box} = 0.54$ \\ 
i\_n100\_z1\_M1 & Inhomogeneous & 100 & 1 & 1 & 27.75 & 0.054 & $15/L_{\rm box} = 0.54$ \\
i\_n100\_z1\_M10 & Inhomogeneous & 100 & 1 & 10 & 27.75 & 0.054 & $15/L_{\rm box} = 0.54$ \\
i\_n100\_z1\_M30 & Inhomogeneous & 100 & 1 & 30 & 27.75 & 0.054 & $15/L_{\rm box} = 0.54$ \\
i\_n100\_z1\_M30\_kmin2.5 & Inhomogeneous & 100 & 1 & 30 & 27.75 & 0.054 & $2.5/L_{\rm box} = 0.09$ \\
i\_n100\_z1\_M30\_kmin5.0 & Inhomogeneous & 100 & 1 & 30 & 27.75 & 0.054 & $5/L_{\rm box} = 0.18$ \\
i\_n1e4\_z1\_M30 & Inhomogeneous & 10$^{4}$ & 1 & 30 & 4 & 0.0078 & $15/L_{\rm box} = 3.75$ \\
\hline
\end{tabular}
\end{center}
\caption{This Table summarizes the parameters of our isolated SNR simulations. The name of the simulation is explanatory: the prefixes `h\_' and `i\_' stand for homogeneous and inhomogeneous medium, respectively. 
The string `nXXX' summarizes the $\bar{n}_{\rm H}$ value in cm$^{-3}$. The string `zXXX' summarizes the metallicity value in solar units. The string `MXX' summarizes the Mach number value which characterizes the density PDF
of the inhomogeneous medium (equations (\ref{eq:lognormal})-(\ref{PDF:3})). For the two runs in which we changed 
the maximum spatial scale of the density fluctuations we also use a string `kminXXX' to indicate the associated minimum wave number. All simulations use AMR with 9 levels of refinement corresponding to $512^3$ resolution 
at maximum refinement.}\label{tab:parameters}
\end{table*}

\subsection{Multiple SNe in a periodic box}
\label{sec:multiple_SNe_descr}
In addition to our simulations of individual SNRs, we also present initial results on the effects of multiple SNRs interacting self-consistently in  a periodic box 
\citep[for previous work in this direction, see][]{2009ApJ...704..137J, 2011ApJ...731...41O, 2011ApJ...743...25K, 2013MNRAS.429.1922C, 2014arXiv1405.7819H}.   Our focus here is on testing the 
sub-resolution SN feedback model that we develop based on our simulations of isolated SNRs (\S \ref{subgrid}). 
In future work, we will extend the present study to quantify in more detail the phase structure of the ISM and how SNe drive turbulence and contribute to galactic winds. 

Our first simulation of multiple SNe is a high resolution simulation designed to resolve the cooling radius (and hence the Sedov-Taylor phase) of all of the SNRs in the box. 
This simulation has a box size of $L_{\rm box}=50$ pc and a Cartesian grid with $256^3$ cells, corresponding to a cell size
$\Delta x =0.195$ pc.  The initial density distribution in the box is homogeneous with $\bar{n}_{\rm H}=100$ cm$^{-3}$ and $Z=Z_{\odot}$. The mean gas density in the simulation box increases slightly with time owing 
to the addition of mass from stellar ejecta, but this is a negligible effect. 
The cooling radius for $\bar{n}_{\rm H} = 100$ cm$^{-3}$ is $\sim 3$ pc and and is thus well resolved at the mean density of the box. 
In practice, once an initial set of SNe explode, the density distribution becomes inhomogeneous and a typical SN explodes in a region of density below the mean density 
of $\bar{n}_{\rm H}=100$ cm$^{-3}$; the cooling radius is then even better resolved (\S \ref{sec:testing_subgrid}).

Every time a SN explodes, $E_{\rm kin} = 9.31 \times 10^{50}$ erg and $E_{\rm th} = 6.9 \times 10^{49}$ erg are injected (as for our isolated SNR simulations; \S \ref{sec:iso_sims_descr}) in a spherical region having a 
radius of 3 cells. The density in the injection region is reset to a constant value equal to the mean density in the 
sphere prior to the explosion (plus a contribution from the ejecta mass).  Momentum and thermal energy are injected cell-by-cell in a volume-weighted fashion, so that momentum conservation is ensured.
We then let SNe shape the properties of the medium at later times. 
The SNe explode at random locations in the box and are not correlated in time. 
We prescribe an average rate $\dot{n}_{\rm SN}=2\times 10^{-4}$ SNe Myr$^{-1}$ pc$^{-3}$ (25 supernovae per Myr for our box). This is typical for massive high redshift star-forming galaxies given their observed star 
formation rates and physical sizes \citep{2010MNRAS.407.2091G}.
Supernovae explode randomly in time, such that the average SN rate is respected on the time scales we consider (a few Myr).
This simulation, which we label `resolved', is more computationally demanding than the others in this paper because 
resolving the Sedov-Taylor phase requires fine time-stepping; we evolve it for 2 Myr, which is of order the turbulent crossing time of the box given the final rms turbulent velocity of $\sim 10$ km s$^{-1}$.

We compare the fully resolved simulation of SN feedback with one in which we implement our sub-resolution SN model (\S \ref{subgrid}).   The SNe explode in a box of the same size as for the resolved case and at the same 
locations and times.  However, the energy and momentum are injected within a sphere of radius 8 pc using the results of the sub-resolution model to determine the partitioning between thermal energy and kinetic energy 
(i.e., radial momentum). These simulations have a lower resolution by a factor of two along each dimension (cell size $\Delta x =0.39$ pc) 
and do not resolve the cooling radius of all SNRs in the box; our goal is to test whether the sub-resolution model based on isolated SNR results captures the correct energy and momentum injected in the turbulent 
medium in that case. 

{The multiple supernova simulations performed in this paper do not reproduce all the properties of the real ISM. As mentioned in the introduction, we in particular neglect spatial and temporal clustering of SNe, 
which can have important effects on the generation of hot bubbles and galactic winds. We also neglect other physical processes (such as ionizing radiation)  that can shape the ambient medium. However, our idealized approach 
allows us carry out a well-defined test of our proposed sub-grid model for SN feedback in a setting in which we can also analytically predict the properties of the resulting turbulence (see \S 6). }

Table~\ref{tab:multiple} summarizes the properties of our multiple SN simulations.

\begin{table*}
\begin{center}
{\bfseries Parameters of multiple supernova simulations}
\end{center}
\begin{center}
\begin{tabular}{lcccccc}
\hline
 Name & $\bar{n}_{\rm H}~[{\rm cm}^{-3}]$ & $Z~[Z_{\odot}]$ & $L_{\rm box}~[{\rm pc}]$ & $\Delta x~[{\rm pc}]$ & $R_{\rm inj}~[{\rm pc}]$ & Resolution \\
\hline
Resolved & 100 & 1 & 50 & 0.195 & 0.59 & 256$^3$ \\
Sub-resolution & 100 & 1 & 50 & 0.39 & 8 & 128$^3$ \\
\hline
\end{tabular}
\end{center}
\caption{Most symbols are the same as in Table \ref{tab:parameters}.  $R_{\rm inj}$ is the radial size of the region in which the thermal energy and momentum of the SNe are injected.  
The `resolved' model deposits a full $10^{51}$ erg (see \S \ref{sec:multiple_SNe_descr}).   The `sub-resolution' model uses the sub-grid model developed in \S \ref{subgrid} in which the division of thermal 
and kinetic energy depends on whether the cooling radius of the SNR is resolved.  Both simulations have SNe rates of  $\dot{n}_{\rm SN}=2\times 10^{-4}$ SNe Myr$^{-1}$ pc$^{-3}$, 
i.e., 25 SNe Myr$^{-1}$ in the box. }\label{tab:multiple}
\end{table*}

\subsection{Comments on neglected physics}

One of the key processes affecting the structure of the inhomogeneous ISM is the interaction between supernova shocks and ambient over-densities in the ISM (``clouds''; e.g., \citealt{Klein1994}).  In the real ISM, 
the mixing of ambient clouds is determined by a combination of thermal conduction, crushing by internal shocks, and fluid instabilities such as the Kelvin-Helmholtz instability. 
The time scale for a shock to shred a cloud of size $\lambda$ due to hydrodynamic instabilities is (e.g., \citealt{Cooper2009})
\begin{equation}
t_{\rm KH}=10^5 \hbox{ yr} \ \kappa \left(\frac{n_{\rm H}/\bar{n}_{\rm H}}{100}\right)^{1/2} \left( \frac{\lambda}{1 \hbox{ pc}} \right) \left( \frac{v_{\rm sh}}{1,000 \hbox{ km/s}} \right)^{-1},
\label{eq:KH}
\end{equation}
where $v_{\rm sh}$ is the speed of the shock relative to the cloud, $n_{\rm H}/\bar{n}_{\rm H}$ is the over-density of the cloud with respect to the mean ambient medium, and $\kappa$ takes into account the importance of 
cooling in the post-shock gas, with $\kappa$ reaching $\sim 10$ for rapid post-shock cooling.   For comparison, the conductive evaporation time in the absence of magnetic fields is (e.g., \citealt{Cowie1977})
\begin{align}
t_{\rm evap} \sim 10^3 \, {\rm yr}  \left(\frac{\lambda}{1 \, {\rm pc}}\right)^{7/6}  \left(\frac{n_{\rm H}/\bar{n}_{\rm H}}{100}\right)^{1/6} \nonumber \\
\times \left(\frac{\bar{n}_{\rm H}}{10 \, {\rm cm^{-3}}}\right)^{1/6}  \left(\frac{T_{\rm hot}}{10^7 \, {\rm K}}\right)^{-5/6} ,
\label{eq:cond}
\end{align}
where $T_{\rm hot}$ is the temperature of the hot medium ablating the cool cloud.
A comparison of equations (\ref{eq:KH}) and (\ref{eq:cond}) suggests that conductive evaporation can be significantly shorter than hydrodynamic mixing in incorporating clouds into the post-shock medium.   
Of course, magnetic fields may significantly modify the conductive evaporation timescale if the field is relatively perpendicular to the temperature gradient.  In addition, equation (\ref{eq:cond}) for the conductive 
evaporation timescale assumes pressure equilibrium between the hot and cool gas.   Since pressure equilibrium is only established on the timescale $t_{KH}/\kappa$ (the ``cloud crushing time'' for the internal shock 
into the cloud to propagate across it), the correct way to interpret this hierarchy of timescales is that the hydrodynamic mixing process that establishes pressure equilibrium is the rate limiting step.   
Once a shock has been driven fully through the cloud, conduction may become quite important in evaporating the cloud, but the former sets the overall timescale for cloud incorporation, not the latter.    
For this reason, we suspect that the neglect of thermal conduction is not a significant limitation in our study.   A more significant uncertainty may be that magnetic fields can alter the mixing of cool and hot 
gas by suppressing Kelvin-Helmholtz and related instabilities, thus modifying the cloud mixing process (e.g., \citealt{Shin2008}, McCourt et al. in prep).   Future work incorporating magnetic fields and anisotropic 
thermal conduction into simulations analogous to those presented here would be very valuable.

{Another caveat is that even our highest resolution simulations may not have enough resolution to faithfully capture the mixing of cool clouds overrun by a SN shocks. In their simulations, \cite{1994ApJ...434L..33M}
found that for radii less than 25 grid cells, the Rayleigh-Taylor instabilities that ought to fragment clouds on a crushing time are suppressed. Nevertheless, we show in the Appendix that our main results appear to be 
well converged. The convergence of our results is aided by the fact that for the power spectra that we study, most of the power is in the largest-scale density fluctuations which are easier to resolve.}

\section{Results for isolated SNRs}
\label{sec:isolated_SNRs}

\subsection{Supernova remnant evolution}
\label{sec:SNR_evolution}

\begin{figure*}
    \includegraphics[width=0.99\textwidth]{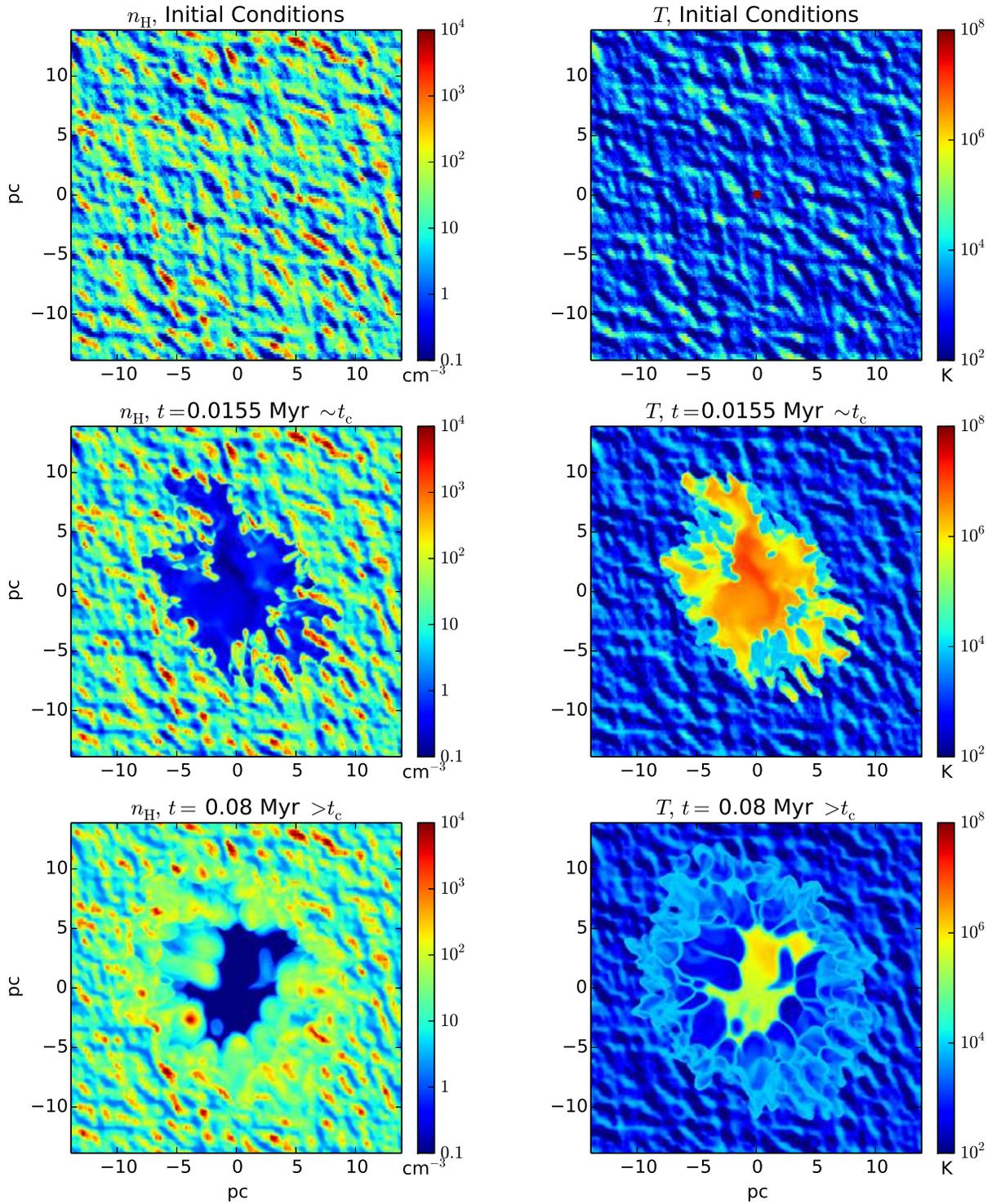}
\caption{ Density (left) and temperature (right) maps for i\_n100\_z1\_M30 during different phases of its evolution. The maps show a slice of thickness 2.75 pc through the centre of the box. For this simulation 
$\bar{n}_{\rm H}=100$ cm$^{-3}$, $\mathcal{M}=30$, $Z=Z_{\odot}$. The top row is the initial condition, the middle row is at roughly the cooling time $t_{\rm c}$, while the bottom row is during the momentum conserving 
phase. Note that the inhomogeneities in the ISM lead to the SNR leaking out preferentially through low density channels.}\label{fig:maps}
\end{figure*}

The evolution of an isolated SNR in an inhomogeneous medium (fiducial run i\_n100\_z1\_M30) is illustrated in Figure~\ref{fig:maps}. 
The top panels show the initial conditions: thermal and kinetic energy injected in a sphere at the centre of the box. 
The middle panels show the effect of the blast wave propagation at time $t=0.015$ Myr, close to the cooling time of a remnant expanding in a homogeneous medium. 
In this phase, the material shocked by the blast wave is still hot ($T>10^6$ K), but a thin layer of cooler gas ($10^4$ K $<T<10^5$ K) develops at the outer edge. 
Due to inhomogeneities in the ambient medium, the evolution of the forward shock and the interaction of 
reverse shocks is complex. 
The remnant is shown well past the cooling time (when it is nearly momentum conserving) in the bottom panels. 
In this late phase, most of the thermal energy has been radiated away, but the blast wave has generated a significant bubble of low-density warm ($T\sim10^{5}-10^{6}$ K) material around the center. 
Large plumes of cooler gas are seen around the over-dense regions that survived mixing with the hot shocked gas. 

\begin{figure*}
  \includegraphics[width=0.99\textwidth]{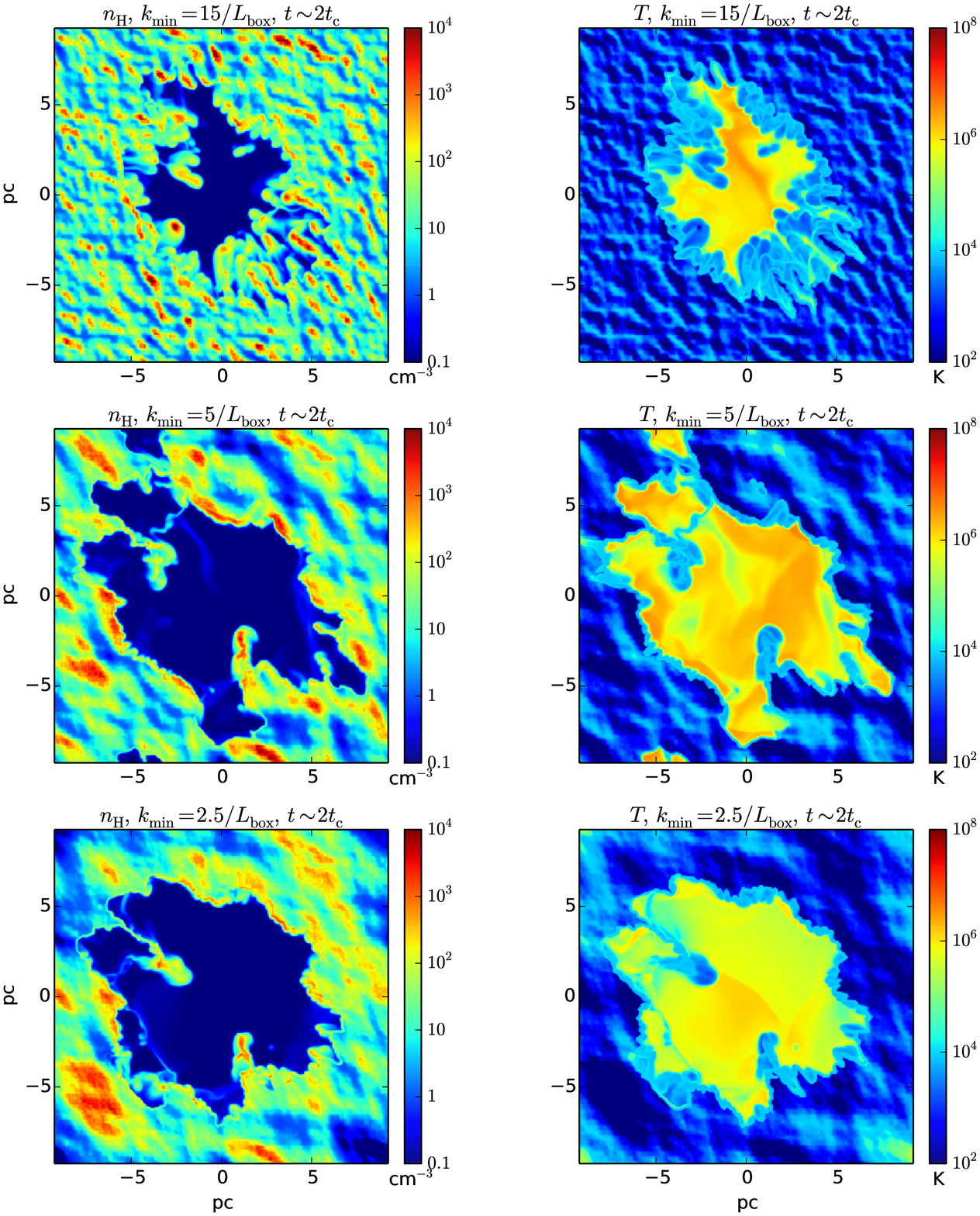} \caption{ Density (left) and temperature (right) maps for runs with different maximum spatial scales in the density power spectrum of the 
  ambient medium, $k_{\rm min}=1/\lambda_{\rm max}$. The maps show a slice of thickness 2.75 pc through the centre of the box at $t = 0.032$ Myr, which is roughly twice the cooling time.  The maximum scale of 
  density fluctuations differs: $k_{\rm min}=15/L_{\rm box}$ (top row), $k_{\rm min}=5/L_{\rm box}$ (mid row), $k_{\rm min}=2.5/L_{\rm box}$ (bottom row); all simulations have 
  $\bar{n}_{\rm H}=100$ cm$^{-3}$, $\mathcal{M}=30$, $Z=Z_{\odot}$. The simulations with lower $k_{\rm min}$ (larger size of density fluctuations) have somewhat more asymmetric evolution.  }\label{fig:mapslc}
\end{figure*}

There are three phenomena that modify the evolution of a radiative blast wave in an inhomogeneous medium relative to the evolution in a homogeneous medium: 
\begin{enumerate}
\item The blast wave propagates faster along paths of least resistance, i.e. along underdense channels.
\item The blast wave shocks over-dense regions as well as underdense regions. 
Overdense regions can be 10 to 100 times denser than the mean ambient density and their post shock temperature can be closer to the peak of the cooling curve than in less dense regions. 
These effects can shorten the local cooling time by a factor 10 to 100 and lead to enhanced radiative losses.
\item ISM clouds mix the gas in the expanding blast wave.
\end{enumerate}

All three of these processes are evident in our simulations.  The first and third are visible in the images shown in Figure~\ref{fig:maps}, where the SN shock clearly propagates preferentially in low density regions 
and over-densities in the ISM initial conditions (top panel) are incorporated and mixed into the SNR as time goes on.     
To partially disentangle the importance of these different processes, Figure~\ref{fig:mapslc} shows density and temperature maps for three runs with different $k_{\rm min}=1/\lambda_{\rm max}$.  
In the initial conditions for these simulations the total variance of the density PDF is fixed but the range of  $\lambda$ varies, implying that each spatial scale possesses a different amount of power in 
the different calculations.   Figure \ref{fig:mapslc} shows that for larger $\lambda_{\rm max}$, the existence of density fluctuations on larger scales results in a more pronounced large-scale asymmetry in the evolution 
of the SNR.   Larger density fluctuations are also shredded more slowly by hydrodynamical instabilities (see equation (\ref{eq:KH})).   

\begin{figure}
  \includegraphics[width=0.5\textwidth]{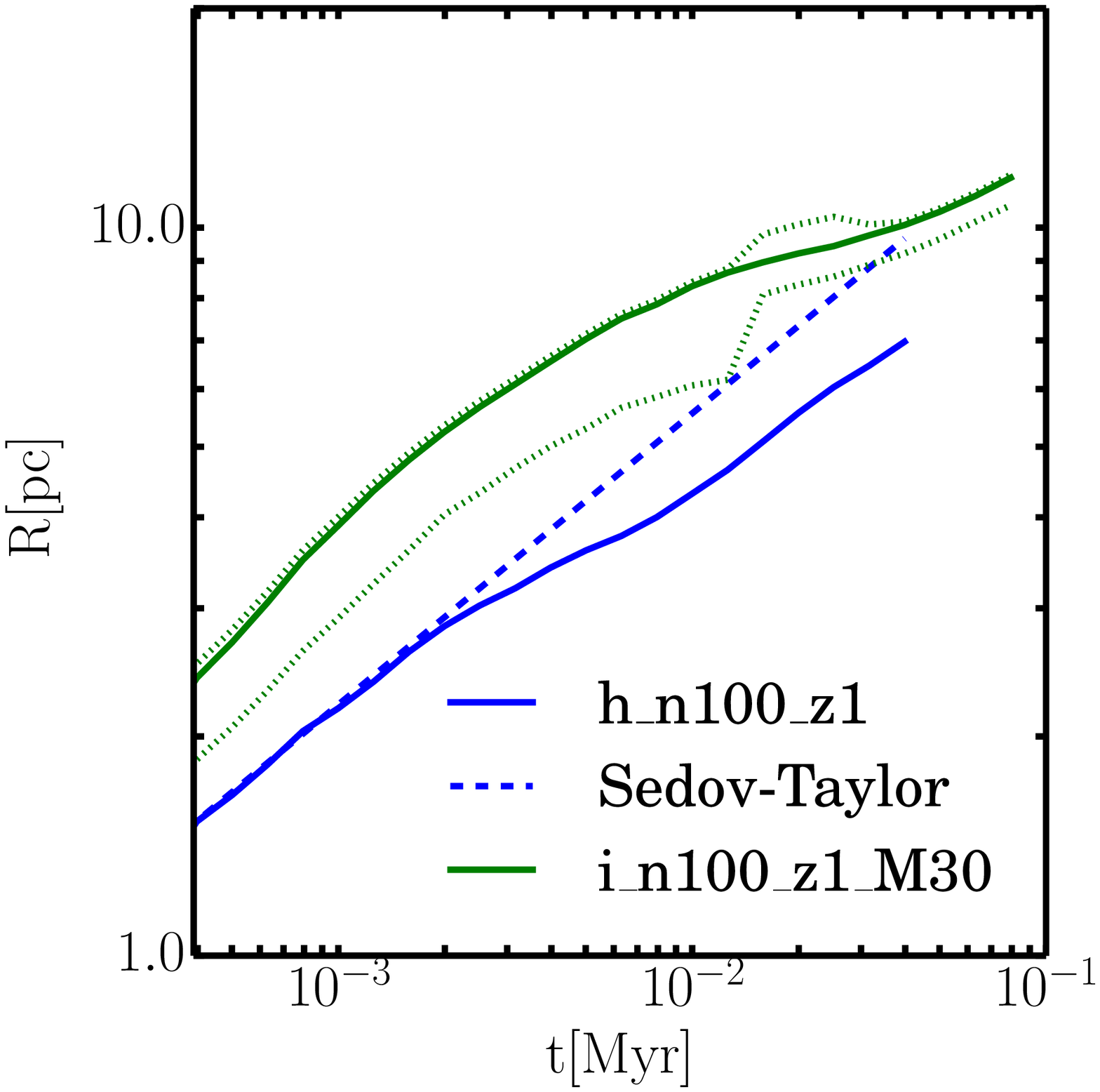} \caption{ Evolution of the forward shock radius $R$ as a function of time $t$ in a homogeneous (blue solid) vs. inhomogeneous 
  (green solid) medium.  The Sedov-Taylor solution is shown with the blue dashed line.  In both simulations the mean density is $\bar{n}_{\rm H}=100$ cm$^{-3}$, the metallicity is $Z=Z_{\odot}$. The Mach number 
  is $\mathcal{M}=30$ for the inhomogeneous medium simulation.  The dotted lines show the scatter obtained when measuring $R(t)$ in 64 equal sections of solid angle.   In the inhomogeneous medium, the SNR propagates more 
  quickly due to the existence of low density channels (see Figures \ref{fig:maps} \& \ref{fig:mapslc}).}\label{fig:radius}
\end{figure}

Figure~\ref{fig:radius} shows the forward shock radius $R$ as a function of time for the homogeneous and inhomogeneous simulations with the same mean density.
For the homogeneous medium case, the shock radius is identified by measuring spherically 
averaged density and temperature profiles and locating jumps in these quantities. 
For the inhomogeneous case, we approximate the outer shock radius $R$ as the radius of the sphere enclosing 99\% of 
the total energy (kinetic + thermal) in the gas (we compared several other methods and this was the most robust).
In the inhomogeneous case, we also determine the radius of the forward shock 
in 64 angular bins of equal solid angle.  The scatter in radius as a function of angle is shown by the dotted lines in Figure~\ref{fig:radius} (the scatter is defined
to be the maximum and minimum radius among the 64 angular bins).

Figure~\ref{fig:radius} shows that during its early evolution ($t<2\times 10^{-3}$ Myr), the expansion of the SNR is well approximated by $R\propto t^{2/5}$, the standard Sedov-Taylor scaling.
The evolution is altered at later times due to radiative cooling. 
For a homogeneous medium, the evolution is well approximated by 
a broken power-law \citep{1988ApJ...334..252C}. 
On average, the blast wave propagates faster in the inhomogeneous case than in the homogeneous case because much of the volume is filled by gas with a density well below the mean density of the box.   
There is also significant scatter in $R(t)$, which depends on the size of ISM clouds (Figure \ref{fig:mapslc}).  Larger clouds are shredded more slowly and generate a more asymmetric remnant, i.e.  larger scatter in $R(t)$.

\begin{figure*}
  \includegraphics[width=0.99\textwidth]{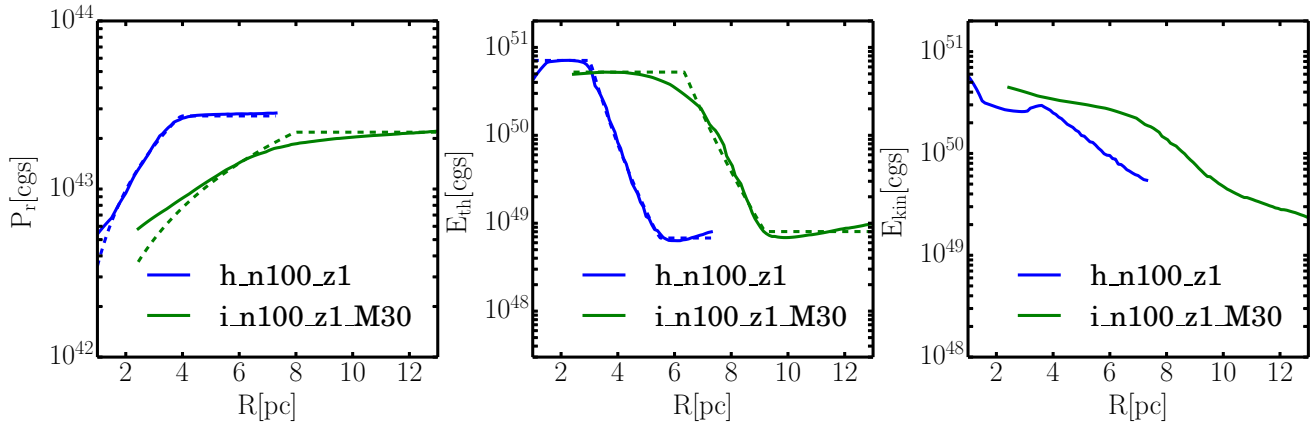} \caption{ Radial momentum $P_{\rm r}$ (left panel), thermal energy $E_{\rm th}$ (middle panel) and kinetic energy $E_{\rm kin}$ (right panel)
  in the SNR as a function of shock position $R$ for a homogeneous (blue) and inhomogeneous (green) ambient ISM. In both simulations the mean density is $\bar{n}_{\rm H}=100$ cm$^{-3}$, the metallicity is $Z=Z_{\odot}$; the 
  Mach number is $\mathcal{M}=30$ for the inhomogeneous ISM. The inhomogeneous medium leads to a larger effective cooling radius because the SNR can propagate rapidly out through low density channels. The final thermal energy 
  is similar in both cases while the final momentum is lower by $\sim 30\%$ in the inhomogeneous medium. The lower radial momentum in the inhomogeneous medium largely accounts for the lower kinetic energy at a given radius.  
  Dashed lines show the fits described in \S~\ref{sec:fits}.}\label{hom_vs_inh}.
\end{figure*}

\subsection{Momentum and thermal energy injection from isolated SNRs}
\label{sec:mom_egy_isolated}
We now quantify the energy and momentum in SNRs as a function of radius and how the evolution varies with the properties of the ISM.  We begin by comparing the evolution in an inhomogeneous ISM to that in a homogeneous 
medium at fixed mean density $\bar{n}_{\rm H}=100$ cm$^{-3}$ (i\_n100\_z1\_M30 and h\_n100\_z1, respectively).  The inhomogeneous medium has density fluctuations set by that expected for isothermal turbulence with Mach 
number $\mathcal{M} = 30$.

\begin{figure*}
  \includegraphics[width=0.99\textwidth]{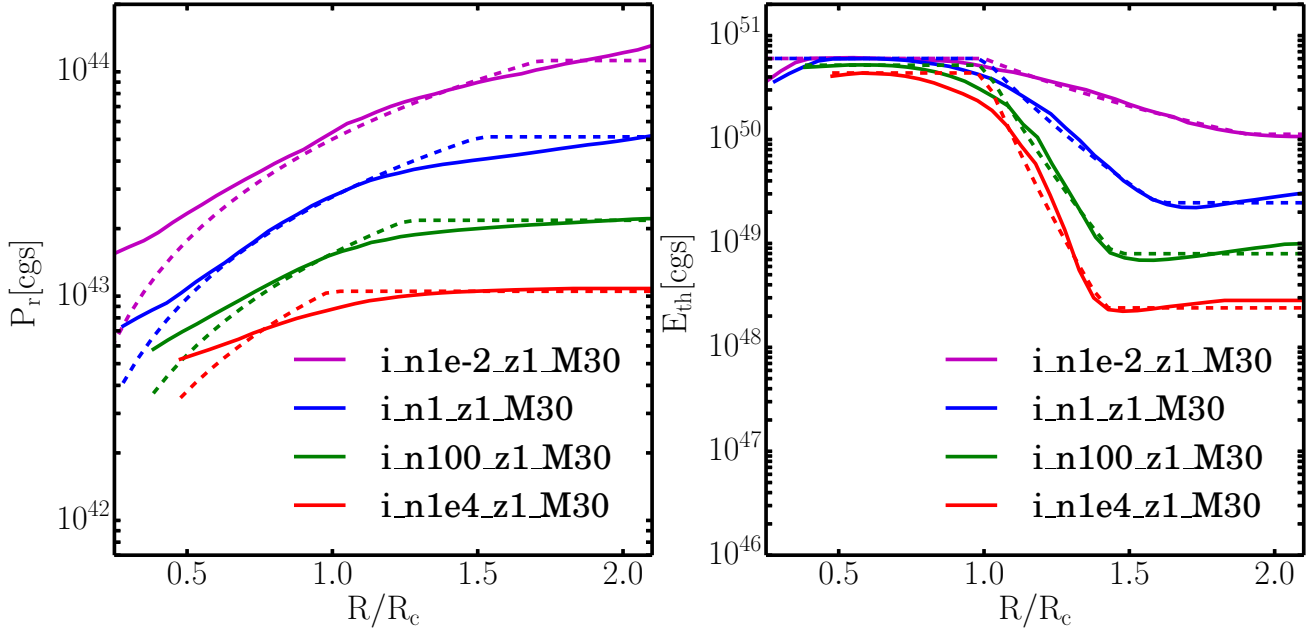} \caption{ Radial momentum $P_{\rm r}$ (left panel) and thermal energy $E_{\rm th}$ (right panel) as a function of shock position $R$ for different mean ambient ISM densities: $ \bar{n}_{\rm H}=10^{-2}$ cm$^{-3}$ (magenta), $ \bar{n}_{\rm H}=1$ cm$^{-3}$ (blue), $\bar{n}_{\rm H}=100$ cm$^{-3}$ (green), $\bar{n}_{\rm H}=10^4$ cm$^{-3}$ (red).  The simulations shown here have Mach number $\mathcal{M}=30$ and metallicity $Z=Z_{\odot}$. Dashed lines show the fits described \S~\ref{sec:fits}. }\label{rhodep}
\end{figure*}

\begin{figure*}
  \includegraphics[width=0.99\textwidth]{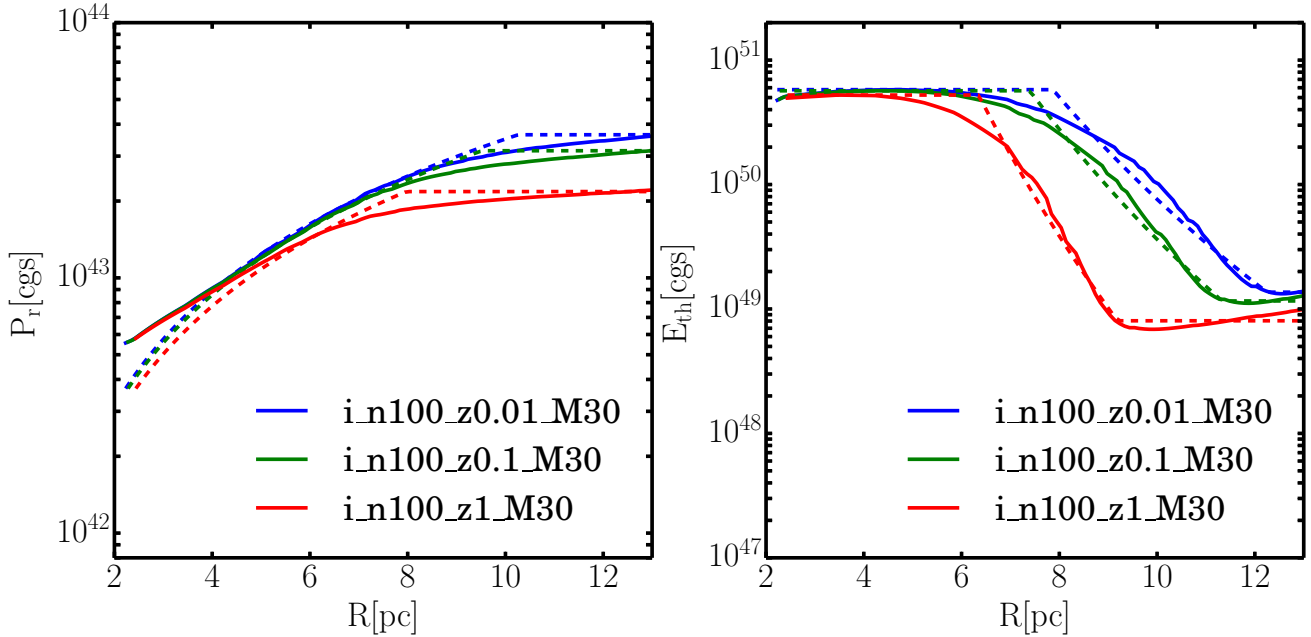} \caption{ Radial momentum $P_{\rm r}$ (left panel) and thermal energy $E_{\rm th}$ (right panel) as a function of forward shock 
  position $R$ for different metallicities of the ISM: $Z=0.01Z_{\odot}$ (blue), $Z=0.1Z_{\odot}$ (green), $Z=Z_{\odot}$ (red).  The simulations all have Mach number $\mathcal{M}=30$ and mean density 
  $\bar{n}_{H}=100$ cm$^{-3}$. Dashed lines show the fits described in \S~\ref{sec:fits}.   }\label{metdep}
\end{figure*}

Figure~\ref{hom_vs_inh} compares the radial momentum $P_{\rm r}$, the thermal energy $E_{\rm th}$, and the kinetic energy $E_{\rm kin}$ as a function of the outer shock radius. The dashed lines in 
Figure \ref{hom_vs_inh} (as well as Figures \ref{rhodep} and \ref{metdep}) are fits described in \S \ref{sec:fits}.

For the homogeneous medium, the solution in Figure \ref{hom_vs_inh} is consistent with the Sedov-Taylor solution with constant thermal and kinetic energy and rising momentum $P_{\rm r}\propto R^{3/2}$ for 
$R\lesssim 3\hbox{ pc}$ (very early on in the SNR evolution, the kinetic energy decreases and the thermal energy increases to approach the approximately constant values associated with the Sedov-Taylor phase).
When the blast wave reaches $R\gtrsim3$ pc (roughly the cooling radius), the thermal and kinetic energy start decreasing owing to radiative losses. For $R\gtrsim 4$ pc, most of the energy has been radiated away and the 
remnant conserves momentum. During the momentum conserving phase at the end of the SNR evolution the kinetic energy decreases because it scales as $E_{\rm kin}\sim P_{\rm r}^2/2M$, where $M$ is the swept up mass, which 
continuously increases.   In future Figures we will omit the kinetic energy panel in describing the SNR evolution because most of the key information conveyed by the kinetic energy is already contained in the radial 
momentum of the remnant, which is also the better-conserved quantity at late times.

The SNR evolution for the inhomogeneous medium in Figure \ref{hom_vs_inh} is qualitatively similar to the homogeneous medium evolution, but quantitatively different.  Early radiative losses related to the presence of 
rapid cooling in overdense regions decreases the energy and momentum of the SNR in the Sedov-Taylor phase.  A plateau in thermal energy is reached, but at a value somewhat lower than in the homogenous medium.  In addition, 
the overall evolution takes longer, with the momentum and thermal energy reaching their asymptotic values only when the shock has reached $\sim 7-10$ pc, rather than $\sim 3$ pc as in the homogeneous medium.
A common result between the homogeneous and inhomogenous medium solutions is that the residual thermal energy at late times is a factor $\sim100$ smaller than the initial input energy. The residual thermal energy is 
associated  with material heated by reverse shocks near the remnant's centre in Figures \ref{fig:maps} and \ref{fig:mapslc}.

Insight into how SNRs evolve in an inhomogeneous medium is provided by considering the evolution of an SNR in a homogeneous medium with a  density representative of the expansion along paths of least resistance 
in the inhomogeneous medium.  Assuming that the ambient medium has a lognormal density PDF as in equation (\ref{eq:lognormal}), \cite{2013MNRAS.433.1970F} give an approximation for the effective density occupied by 
most of the volume $n_{\rm H,eff}$, defined such that the volume fraction of gas with density below $n_{\rm H,eff}$ is 50\%: 
\begin{equation} 
n_{\rm H,eff} \approx 0.06 \left( \frac{\mathcal{M}}{30}\right)^{-1.2} \bar{n}_{\rm H}.
\end{equation}
Let us assume that the cooling radius of SNRs scales with density as $R_{\rm c}\propto n_{\rm H}^{-3/7}$ \citep{1988ApJ...334..252C}.    In this case the cooling radius in the inhomogeneous medium will be a factor 
of $\sim 3$ larger for $\mathcal{M} = 30$.
Figure \ref{hom_vs_inh} suggests that the cooling radius increases by closer to a factor of $2$, somewhat less than suggested by this simple argument.  Part of the reason for this is that the cooling is $\propto n^2$ and is 
thus sensitive to the denser phases of the inhomogeneous ISM, not simply to $n_{\rm H,eff}$.

One of the most important drivers of ISM turbulence is the residual late time momentum of SNRs.   This significantly exceeds the input momentum in the initial ejecta because of work done on the swept up mass during the 
Sedov-Taylor phase. We define the initial radial momentum of the remnant (immediately after the stellar explosion) as
\begin{equation}
P_0=\left(2M_{\rm ej}E_{\rm 0}\right)^{1/2},
\end{equation}
where $M_{\rm ej}$ is the ejecta mass and $E_{\rm 0}$ is the initial (total) energy of the SN.  We also define $P_{\rm fin}$ as the asymptotic radial momentum obtained after most of the thermal energy has been radiated away.
For our fiducial homogeneous and inhomogeneous medium simulations (h\_n100\_z1 and i\_n100\_z1\_M30), we find $P_{\rm fin}/P_0\approx 8.1$ and $P_{\rm fin}/P_0\approx 6.0$, respectively.   
Thus, the momentum boost is  $\sim30\%$ lower in the inhomogeneous medium. This decrease in the final momentum is primarily due to the presence of high density inhomogeneities which have  short post-shock cooling 
times and cause significant early radiative energy loss.    Note that the inhomogeneous ISM simultaneously increases the cooling radius (due to low density channels that the shocked gas can propagate through) and 
decreases the final momentum (due to high density regions that enhance cooling).

Figure~\ref{rhodep} shows how the radial momentum and thermal energy of SNRs
as a function of outer shock radius varies with the mean density of the ambient medium. 
To show the results of simulations with different mean densities on the same plot, the radius on the horizontal axis has been scaled relative to the cooling radius $R_{\rm c}$ (quantified in the next subsection for 
our inhomogeneous ISM simulations). 
In agreement with analytic expectations \citep[e.g.,][]{1988ApJ...334..252C}, the asymptotic radial momentum decreases with increasing ambient medium density. 
Since cooling is more efficient in high density environments, the residual thermal energy at late times also decreases with density. It turns out that the final kinetic energy $E_{\rm kin}\sim P_{\rm r}^2/2M$ is always larger 
than the final thermal energy by a factor $\sim 2-3$ (depending on the mean density of the ISM): supernovae mostly inject kinetic energy (i.e. momentum) at scales larger than the cooling radius. 
The radial momentum in simulation i\_n1\_z1\_M30 and i\_n1e-2\_z1\_M30 (the lowest density cases) has not fully converged at the end of the simulation (which we stopped when the shock reached the edge of the computational box). 
For i\_n1\_z1\_M30 ($ \bar{n}_{\rm H}=1$ cm$^{-3}$) the momentum is not expected to increase significantly beyond this point (e.g., in the simulations of \cite{1988ApJ...334..252C}, the momentum does not increase by more than 
20\% beyond twice the cooling radius). {For the lowest ambient density simulation i\_n1e-2\_z1\_M30 ($ \bar{n}_{\rm H}=10^{-2}$ cm$^{-3}$), the cooling radius inferred from the decline in thermal energy is $\sim 300$ pc (almost twice 
the value in a homogeneous medium); however when $R\sim 100$ pc the RMS velocity of the gas in the remnant is already $<10$ km/s and the SNR would merge with the turbulent ISM before reaching the snowplow phase. This effect is not 
fully captured in our simulation because we do not adopt a self-consistent model for the velocity structure of the turbulent ISM.}

Since the cooling function depends on metallicity, the evolution of SNRs also depends on the metallicity of the ambient medium, as shown in Figure~\ref{metdep}. 
Metal line emission increases the cooling rate of the gas and thus results in lower final momentum and thermal energy. 
For $Z<0.01Z_{\odot}$, metal lines contribute only a small fraction of the total cooling rate and the results therefore do not depend sensitively on metallicity. 
Partial convergence of the final momentum and thermal energy is already seen when comparing the results for $Z=0.1Z_{\odot}$ and $Z=0.01Z_{\odot}$. 

\begin{figure*}
    \includegraphics[width=0.99\textwidth]{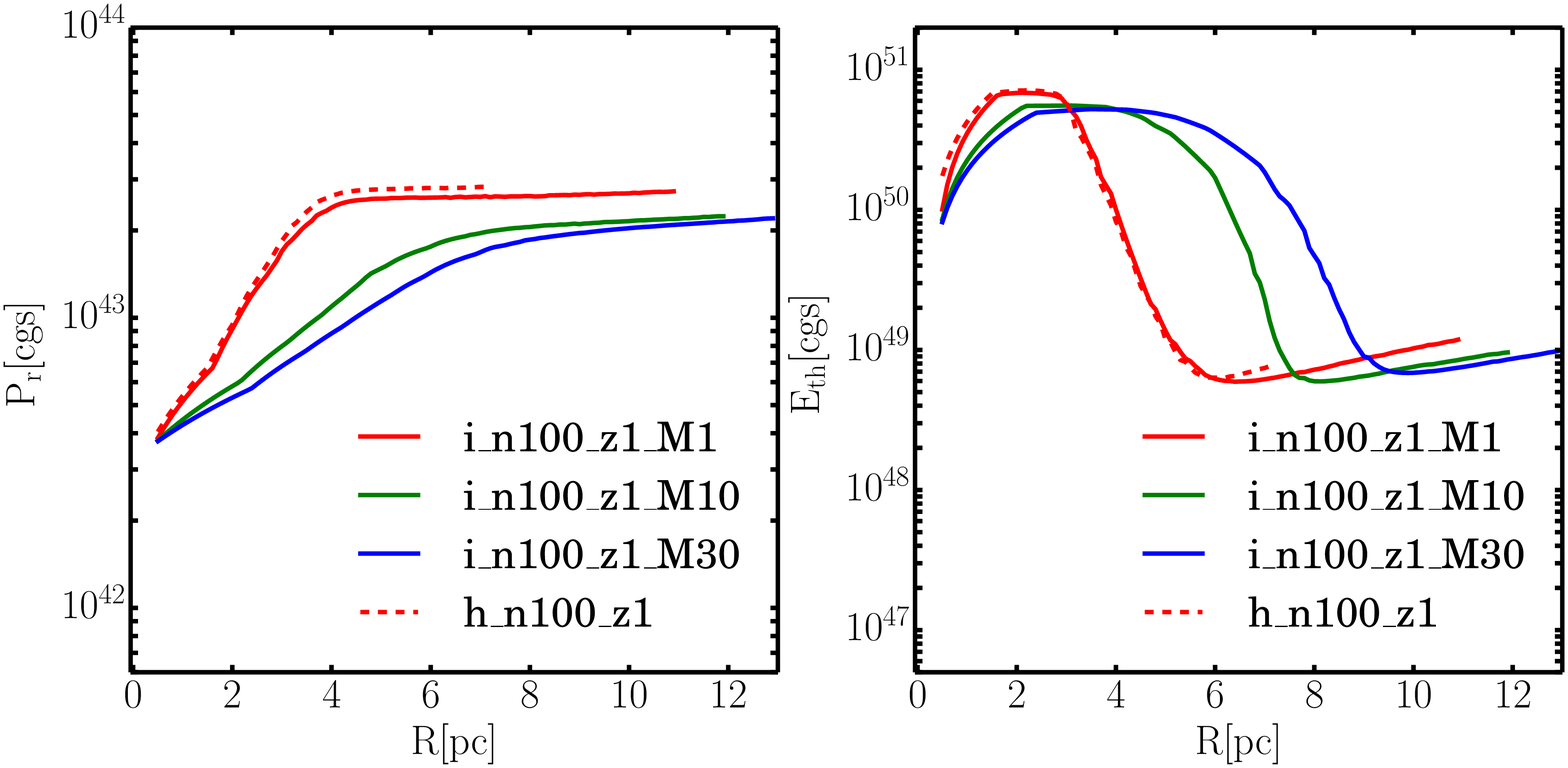}
\caption{ Radial momentum $P_{\rm r}$ (left panel) and thermal energy $E_{\rm th}$ (right panel) as a function of shock position $R$ for different ambient medium density fluctuations, characterized by the Mach number 
of the lognormal density PDF:  $\mathcal{M}=30$ (solid blue), $\mathcal{M}=10$ (solid green), $\mathcal{M}=1$ (solid red). Mean density $\bar{n}_{\rm H}=100$ cm$^{-3}$ and metallicity $Z=Z_{\odot}$ are held fixed.  
The case for the homogeneous medium is shown as a red dashed line. In the supersonic limit, the results depend weakly on Mach number while the results converge to the homogeneous medium case 
for $\mathcal{M}\le 1$.}\label{machdep}
\end{figure*}

We have also explored how the SNR evolution is affected by the density structure of the ambient medium, which is parameterized by the Mach number $\mathcal{M}$ of the turbulence and the maximum spatial scale of the 
density fluctuations,  $\lambda_{\rm max}$.
Figure~\ref{machdep} shows the results of varying the Mach number  while holding the other parameters fixed.
For $\mathcal{M}> 10$, the final momentum depends only weakly on Mach number, implying that a sub-resolution model for momentum injection by SNe does not require \emph{a priori} knowledge of the Mach number in this limit. 
For $\mathcal{M}< 10$, the solution approaches the homogeneous medium case, as expected on physical grounds. 
The evolution of the thermal energy is a strong function of Mach number, with higher Mach numbers corresponding to the ISM having more low density channels and thus the SNR having a larger effective cooling radius.    
However,  the final thermal energy obtained at large radii is the same within a factor $\approx2$ for $\mathcal{M}=1-30$.   

\begin{figure*}
  \includegraphics[width=0.99\textwidth]{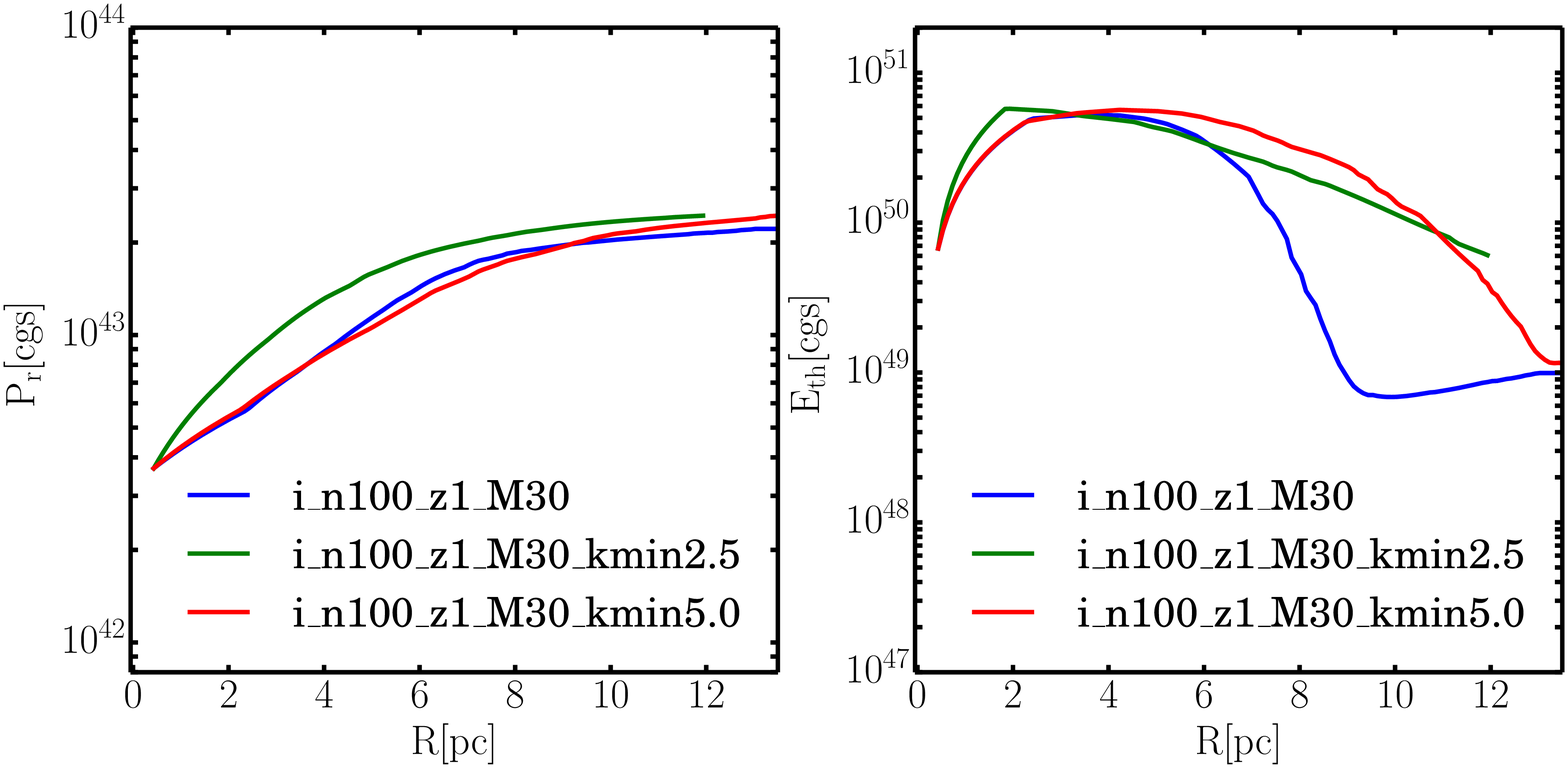} \caption{ Radial momentum $P_{\rm r}$ (left panel) and thermal energy $E_{\rm th}$ (right panel) as a function of shock radius $R$ for 
  different values of the maximum spatial scale of the density fluctuations in the ambient ISM: $k_{\rm min}=15/L_{\rm box}$ (blue), $k_{\rm min}=5/L_{\rm box}$ (green), $k_{\rm min}=2.5/L_{\rm box}$.  
  The final radial momentum does not depend significantly on clump size but the evolution of the thermal energy does. Mean density $\bar{n}_{\rm H}=100$ cm$^{-3}$, metallicity $Z=Z_{\odot}$ and Mach number 
  $\mathcal{M}=30$ are held fixed.}\label{clumpdep}
\end{figure*}

Figure~\ref{clumpdep} shows how the evolution of SNRs varies as we change $\lambda_{\rm max}$, the maximum spatial scale of the density fluctuations in the ambient medium (the corresponding images are shown in 
Figure~\ref{fig:mapslc}).
Larger $\lambda_{\rm max}$ corresponds to low density regions that are more spatially coherent, and to larger stochastic effects since $\lambda_{\rm max}$ can be comparable to, or larger than, the cooling radius. In fact, 
if we consider a sphere of radius 4 pc (roughly the radius at which the momentum of the remnant reaches the asymptotic value in the homogeneous medium), we measure an average density $<n>=98.6$ cm$^{-3}$ for 
$\lambda_{\rm max}=L_{\rm box}/15$ which is close to the mean in the box. However, we measure different densities for larger $\lambda_{\rm max}$: $<n>=68.3$ cm$^{-3}$ for 
$\lambda_{\rm max}=L_{\rm box}/5$ and $<n>=90.6$ cm$^{-3}$ for $\lambda_{\rm max}=L_{\rm box}/2.5$. The somewhat non-monotonic behaviour with $\lambda_{\rm max}$ in Figure~\ref{clumpdep} is partially due to this difference in 
the mean density in the vicinity of the SN.
However, once a large volume has been swept up, our simulations show that the radial momentum converges to approximately the same value for the three different $\lambda_{\rm max}$.  
Varying $\lambda_{\rm max}$ has a much stronger effect on the thermal energy evolution. 
For larger $\lambda_{\rm max}$, there are more large clouds which take a longer time to mix (equation (\ref{eq:KH})) and radiate away their energy, resulting in a slower decrease of the thermal energy in the remnant. 

\section{Fits to the simulation results}
\label{sec:fits}

The simulation results described in \S \ref{sec:isolated_SNRs} can be fit using simple functions of the shock radius $R$.  These fits are plotted as dashed lines in Figures~\ref{hom_vs_inh}, \ref{rhodep} and \ref{metdep} 
and do a good job at reproducing the basic features of the simulations. We adopt the following fitting formulae for the momentum and thermal energy associated with a single SN:
\begin{eqnarray}\label{formula1}
E_{\rm th}(R)&=& E_{\rm th,0}\theta(R_{\rm c}-R) \nonumber \\
&+&E_{\rm th,0}\left(\frac{R}{R_{\rm c}}\right)^{\alpha}\theta[(R-R_{\rm c})(R_{\rm r}-R)] \nonumber \\
&+&E_{\rm th,0}\left(\frac{R_{\rm r}}{R_{\rm c}}\right)^{\alpha}\theta(R-R_{\rm r}) 
\end{eqnarray}
\begin{eqnarray}\label{formula2}
P_{\rm r}(R)&=& P_0\left(\frac{R}{R_0}\right)^{1.5}\theta(R_{\rm b}-R)\nonumber \\
&+&P_0\left(\frac{R_{\rm b}}{R_0}\right)^{1.5}\theta(R-R_{\rm b})
\end{eqnarray}
where $\theta(x)$ is the Heaviside step function. The parameters of these fitting functions are as follows: $E_{\rm th,0}$ is the thermal energy during the Sedov-Taylor phase (a constant determined by the initial conditions); 
$R_{\rm c}$ is the radius beyond which cooling 
becomes important (i.e., the cooling radius); $\alpha$ is a slope characterizing the decline in thermal energy past $R_{\rm c}$; $R_{\rm r}$ represents the late-time radius beyond which thermal energy is approximately 
constant or slowly rising;\footnote{The subscript {\itshape r} stands for \textquotedblleft rise\textquotedblright.}
$P_0$ is the initial momentum, which is set by the initial conditions; $R_0$ is a scale radius used to extrapolate the Sedov-Taylor power law to small radii; the slope of $P_{\rm r}(R)$ during the Sedov-Taylor phase in the 
first term in equation (\ref{formula2}) is fixed to a value of 1.5 which is appropriate for the analytical solution in a homogeneous medium; and $R_{\rm b}$ is the radius beyond which the radial momentum of the SNR is 
approximately constant $P_{\rm r}(R)=P_0(R_{\rm b}/R_0)^{1.5}=P_{\rm fin}$.\footnote{The subscript {\itshape b} stands for \textquotedblleft break.\textquotedblright}

Equations (\ref{formula1}) \& (\ref{formula2}) are simple functions that accurately describe the final stage of SNR evolution and roughly reproduce the evolution in the early phase (as shown by the comparison to the 
full simulation results in Figures \ref{hom_vs_inh}-\ref{metdep}).  However, they do not describe the evolution prior to the onset of the Sedov-Taylor phase nor are they perfect fits in the phase during which the thermal 
energy declines rapidly. These formulae are somewhat more accurate for simulations with a homogeneous medium because in this case the analytical solution is roughly a piece-wise power law \citep{1988ApJ...334..252C}.

We fit equations (\ref{formula1}) \& (\ref{formula2}) to each of our simulations. In doing so, we assume a precision of 5\% in measuring the momentum and thermal energy in the simulations (i.e., 5\% error bar weights in a 
standard $\chi^2$ fit). All data points are equally spaced in time. The values of the resulting parameters are summarized in Table~\ref{tab:fitparameters}.   For the inhomogeneous medium fits, we consider only the fiducial 
case with $\mathcal{M}=30$. Using the fiducial parameters for i\_n100\_z1\_M30 and h\_n100\_z1 as pivot points, we fit for the scalings of the fitting parameters with density and metallicity. 
Power-laws are fit for metallicity and density separately. For all the fits below (equations (11)-(19)), the results for $Z\lesssim 0.01 Z_{\odot}$ should be obtained by simply using the values at $Z= 0.01 Z_{\odot}$ rather 
than extrapolating to even lower $Z$. This is because the cooling becomes relatively independent of metallicity at low $Z$.

For the {\em homogeneous medium}, the fitting parameters scale with density and metallicity as follows:
\begin{eqnarray}
&\alpha=&-7.8\left(\frac{Z}{Z_{\odot}}\right)^{0.050}\left(\frac{n_{\rm H}}{100 {\rm~cm}^{-3}}\right)^{0.030} \nonumber \\
&R_{\rm c}=&3.0 \hbox{\rm~pc} \left(\frac{Z}{Z_{\odot}}\right)^{-0.082}\left(\frac{n_{\rm H}}{100 {\rm~cm}^{-3}}\right)^{-0.42} \nonumber \\
&R_{\rm r}=&5.5 \hbox{\rm~pc} \left(\frac{Z}{Z_{\odot}}\right)^{-0.074}\left(\frac{n_{\rm H}}{100 {\rm~cm}^{-3}}\right)^{-0.40} \nonumber \\
&R_{0}=&0.97 \hbox{\rm~pc} \left(\frac{Z}{Z_{\odot}}\right)^{0.046}\left(\frac{n_{\rm H}}{100 {\rm~cm}^{-3}}\right)^{-0.33} \nonumber \\
&R_{\rm b}=&4.0 \hbox{\rm~pc} \left(\frac{Z}{Z_{\odot}}\right)^{-0.077}\left(\frac{n_{\rm H}}{100 {\rm~cm}^{-3}}\right)^{-0.43}. 
\label{eq:fits-h}
\end{eqnarray}
The scalings are slightly different for the {\em inhomogeneous medium} with 
$\mathcal{M}=30$ and $\lambda_{\rm max} = L_{\rm box}/15$:
\begin{eqnarray}
&\alpha=&-11\left(\frac{Z}{Z_{\odot}}\right)^{0.070}\left(\frac{n_{\rm H}}{100 {\rm~cm}^{-3}}\right)^{0.114} \nonumber \\
&R_{\rm c}=&6.3 \hbox{\rm~pc} \left(\frac{Z}{Z_{\odot}}\right)^{-0.050}\left(\frac{n_{\rm H}}{100 {\rm~cm}^{-3}}\right)^{-0.42} \nonumber \\
&R_{\rm r}=&9.2 \hbox{\rm~pc} \left(\frac{Z}{Z_{\odot}}\right)^{-0.067}\left(\frac{n_{\rm H}}{100 {\rm~cm}^{-3}}\right)^{-0.44} \nonumber \\
&R_{0}=&2.4 \hbox{\rm~pc} \left(\frac{Z}{Z_{\odot}}\right)^{0.021}\left(\frac{n_{\rm H}}{100 {\rm~cm}^{-3}}\right)^{-0.35} \nonumber \\
&R_{\rm b}=&8.0 \hbox{\rm~pc} \left(\frac{Z}{Z_{\odot}}\right)^{-0.058}\left(\frac{n_{\rm H}}{100 {\rm~cm}^{-3}}\right)^{-0.46}.
\label{eq:fits-i}
\end{eqnarray}

We also fit for the evolution of the shock position as a function of time $R(t)$. This function can be easily used to convert the expressions for $P_{\rm r}(R)$ and $E_{\rm th}(R)$ to functions of time $t$. 
For the purpose of the fits, we define the cooling time $t_{\rm c}$ as the time at which $R(t_{\rm c})=R_{\rm c}$ is reached. Since the early evolution of the SNR is well described by a Sedov-Taylor solution, we assume 
the analytical solution $R\propto t^{2/5}$ for $t<t_{\rm c}$ \citep{1988ApJ...334..252C}. However, in our isolated SNR simulations we find that for $t\gg t_{\rm c}$ the slope is slightly different than the one expected for 
the pressure-driven snowplow ($R\propto t^{2/7}$) and for the momentum-conserving phase ($R\propto t^{1/4}$), so we generalize the fitting function using a power-law scaling $R\propto t^{\beta}$ where $\beta$ is a parameter 
of the fits. We find that $\beta$ does not strongly depend on density and metallicity (within the assumed uncertainty). For the {\em homogeneous medium} we find: 
\begin{equation}
 R(t)=R_{\rm c}\left(\frac{t}{t_{\rm c}}\right)^{2/5}\theta(R_{\rm c}-R)+R_{\rm c}\left(\frac{t}{t_{\rm c}}\right)^{0.29},
\end{equation}
where the cooling time is
\begin{equation}
t_{\rm c}=2,400\hbox{~yr} \left(\frac{n_{\rm H}}{100 \hbox{~cm}^{-3}}\right)^{-0.54} \left(\frac{Z}{Z_{\odot}}\right)^{-0.20}
\label{tc-h}
\end{equation}
and the cooling radius $R_c$ is given in equation (\ref{eq:fits-h}).   For the {\em inhomogeneous medium} we find:
\begin{equation}
 R(t)=R_{\rm c}\left(\frac{t}{t_{\rm c}}\right)^{2/5}\theta(R_{\rm c}-R)+R_{\rm c}\left(\frac{t}{t_{\rm c}}\right)^{0.20},
\end{equation}
where the cooling time is
\begin{equation}
t_{\rm c}=3,500\hbox{~yr} \left(\frac{n_{\rm H}}{100 \hbox{~cm}^{-3}}\right)^{-0.53} \left(\frac{Z}{Z_{\odot}}\right)^{-0.17}
\label{tc-i}
\end{equation}
and the cooling radius $R_c$ is given in equation (\ref{eq:fits-i}).  Note that the cooling time in equations (\ref{tc-h}) and (\ref{tc-i}) refers to the onset of radiative cooling at the end of the Sedov-Taylor phase.  
The remnant expands by another factor of $\sim 50\%$ in radius ($\lesssim 10$ in time) during the phase over which most of the thermal energy is radiated away (e.g., Figure \ref{hom_vs_inh}).

\begin{table*}
\begin{center}
{\bfseries Parameters of the fitting formulae for SNR evolution}
\end{center}
\begin{center}
\begin{tabular}{lcccccc}
\hline
 Type & $\alpha$ & $R_{\rm c}~[{\rm pc}]$ & $R_{\rm r}~[{\rm pc}]$ & $R_{\rm 0}~[{\rm pc}]$ & $R_{\rm b}~[{\rm pc}]$ & $t_{\rm c}~[{\rm yr}]$ \\
\hline
 h\_n1e4\_z1 & -8.4 & 0.42 & 0.84 & 0.21 & 0.54 & 1.9 $\times 10^2$ \\
 h\_n100\_z0.01 & -6.2 & 4.3 & 7.6 & 0.95 & 5.8 & 6.0 $\times 10^3$ \\
 h\_n100\_z0.1 & -7.0 & 3.8 & 6.7 & 0.96 & 5.0 & 4.6 $\times 10^3$ \\
 h\_n100\_z1 & -7.8 & 3.0 & 5.5 & 0.97 & 4.0 & 2.4 $\times 10^3$ \\
 h\_n1\_z1 & -6.8 & 21 & 34 & 4.4 & 28 & 3.0 $\times 10^4$ \\ 
 i\_n1e4\_z1\_M30 & -15 & 0.91 & 1.3 & 0.43 & 0.90 & 2.9 $\times 10^2$ \\
 i\_n100\_z0.01\_M30 & -8.5 & 7.9 & 12 & 2.2 & 10 & 7.3 $\times 10^3$ \\
 i\_n100\_z0.1\_M30 & -9.1 & 7.4 & 11 & 2.3 & 9.5 & 5.9 $\times 10^3$ \\
 i\_n100\_z1\_M30 & -11 & 6.3 & 9.2 & 2.4 & 8.0 & 3.5 $\times 10^3$ \\
 i\_n1\_z1\_M30 & -6.8 & 43 & 69 & 12 & 66 & 4.0 $\times 10^4$\\
 \hline
\end{tabular}
\end{center}
\caption{ Best fit parameters for the formulae in equations~(\ref{formula1}) and (\ref{formula2}) that describe the momentum and thermal energy of the SNR as a function of radius.  For the inhomogeneous medium, we only 
present results for the fiducial case with $\mathcal{M}=30$ and $k_{\rm min} = 15/L_{\rm box}$.}\label{tab:fitparameters}
\end{table*}

The efficacy of SNe in driving turbulence in the ISM depends on the final momentum of the SNR.  In the context of an ensemble of SNe from a stellar population, the relevant quantity is the momentum injected per 
stellar mass formed $P_{\rm fin}/m_*$.  This quantity (or others like it) is used in analytic models of star formation regulation by stellar 
feedback \citep[e.g.,][]{2005ApJ...630..167T,2011ApJ...731...41O, 2013MNRAS.433.1970F}.  To evaluate this quantity, we assume that one SN is produced for each 100 M$_{\odot}$ of stars formed.  
Table~\ref{tab:momentum} compiles $P_{\rm fin}/P_{0}$ and $P_{\rm fin}/m_{*}$ values for many of our simulations.
\begin{table}
\begin{center}
{\bfseries Asymptotic momentum for isolated SNRs}
\end{center}
\begin{center}
\begin{tabular}{lcc}
\hline
 Type & $P_{\rm fin}/P_0$ & $P_{\rm fin}/m_*$ [km/s] \\
\hline
 h\_n1e4\_z1 & 4.1 & 680 \\
 h\_n100\_z0.01 & 14.9 & 2,590 \\
 h\_n100\_z0.1 & 11.8 & 2,050 \\
 h\_n100\_z1 & 8.1 & 1,420 \\
 h\_n1\_z1 & 15.6 & 2,960 \\
 h\_n1e-2\_z1 & 26.8 & 5,500 \\
 i\_n1e4\_z1\_M30 & 3.1 & 540 \\
 i\_n100\_z0.01\_M30 & 9.8 & 1,820 \\
 i\_n100\_z0.1\_M30 & 8.6 & 1,580 \\
 i\_n100\_z1\_M30 & 6.0 & 1,110 \\
 i\_n1\_z1\_M30 & 11.7 & 2,680 \\
 i\_n1e-2\_z1\_M30 & 26.8 & 5,500 \\
 \hline
\end{tabular}
\end{center}
\caption{ 
$P_{\rm fin}/P_0$ is the ratio of the asymptotic radial momentum to the momentum injected in the SN; $P_{\rm fin}/m_*$ is the asymptotic momentum per unit stellar mass formed, assuming 1 SN per 100 M$_{\odot}$ of 
stars formed.  The inhomogeneous medium simulations correspond to $\mathcal{M}=30$ and $k_{\rm min} = 15/L_{\rm box}$. }
\label{tab:momentum}
\end{table}
We also provide simple scaling relations for how the momentum injected per unit stellar mass formed scales with ambient density and metallicity.
For the homogeneous medium we find:
\begin{equation}
\frac{P_{\rm fin}}{m_{*}}=1,420~\hbox{\rm km s$^{-1}$} \left(\frac{Z}{Z_{\odot}}\right)^{-0.137}\left(\frac{n_{\rm H}}{100~{\rm cm}^{-3}}\right)^{-0.160}
\end{equation}
while for the inhomogeneous medium, we obtain
\begin{equation}
\label{Pfin inhomo}
\frac{P_{\rm fin}}{m_{*}}=1,110~\hbox{\rm km s$^{-1}$} \left(\frac{Z}{Z_{\odot}}\right)^{-0.114}\left(\frac{n_{\rm H}}{100~{\rm cm}^{-3}}\right)^{-0.190}.
\end{equation}

\section{A sub-resolution model for SN feedback}\label{subgrid}
\label{sec:subgrid}
We now use the results of our isolated SN simulations to construct a model for SN feedback for use in simulations that do not have sufficient resolution to resolve the Sedov-Taylor phase during 
which the momentum of the remnant is enhanced and the thermal energy is lost, respectively.
One way to phrase the resolution needed to capture this dynamics is in terms of a {\em mass resolution} requirement to explicitly resolve the momentum boost of SNRs associated with the Sedov-Taylor phase.
The momentum boost is $P_{\rm fin}/P_{\rm 0} \sim (M_{\rm swept}/M_{\rm ej})^{1/2}$. 
For $P_{\rm fin}/P_{\rm 0}\sim 10$ and $M_{\rm ej} = 3$ M$_{\odot}$, this implies that the simulations must resolve a swept up mass $M_{\rm swept}\sim 300$ M$_{\odot}$. 
Zoom-in cosmological simulations typically reach this mass resolution only for 
dwarf galaxies \citep[e.g.,][]{2008ApJ...685...40W, 2014MNRAS.445..581H}.   Even non-cosmological galaxy-scale simulations do not necessarily reach the appropriate resolution.

The fitting formulae summarized in \S~\ref{sec:fits} can be used to inject the proper radial momentum and thermal energy.  A useful rule of thumb is: {\itshape If the injection region is larger than the cooling 
radius given the local ISM properties (density, metallicity, etc.), SN feedback should be implemented primarily via radial momentum deposition; thermal energy injection is sub-dominant $\ll 10^{51}$ erg.  
By contrast, if the injection 
region is smaller than the cooling radius, injecting primarily thermal energy $\sim 10^{51}$ erg is reasonable.  And in an inhomogeneous medium the cooling radius is a factor of few larger than in a homogeneous 
medium with the same 
average density, but the final thermal energy and momentum of the remnant are similar.}

More specifically, given an injection region of size $R$ we advocate using equations (\ref{formula1}) and (\ref{formula2}) to determine the appropriate thermal energy and momentum, respectively, to input into the ISM.  
We suspect that our inhomogeneous medium fits (equations (\ref{eq:fits-i})) are probably a better approximation to the dynamics of the real ISM than the homogeneous medium fits (equation \ref{eq:fits-h}) given the 
likely presence of a wide range of substructure that is not captured in numerical simulations (both because of finite resolution and neglected physics).
One of the important caveats to bear in mind in building such a sub-resolution model is that the thermal energy as a function of radius depends somewhat sensitively on the density distribution of the ambient medium 
(e.g., Figures \ref{machdep} \& \ref{clumpdep}).  This is very difficult to faithfully capture using a sub-resolution model.  Fortunately, however, the asymptotic thermal energy and momentum are much less sensitive to the 
detailed structure of the ISM. 

{It is important to bear in mind that our subgrid model is based on simulations of individual SNe exploding in an inhomogeneous ISM. These simulations do not capture several important effects. At low ambient gas densities, 
SNRs are likely to come into pressure balance (merge) with the ISM before reaching the snowplow phase. In this case, 
the fitting formulae are not valid past the merging radius. For very low ambient densities,  (e.g., $\bar{n}_{\rm H}\sim 10^{-2}$ cm$^{-3}$), the merging radius can be substantially smaller than the cooling radius, in which case the 
momentum injected in the ISM by the SNR can be correspondingly less than the asymptotic value predicted by our fitting formulae. At low densities, SNRs are also more likely to merge with one another and generate hot bubbles 
that break out of the galactic disc before reaching the cooling radius. In that case, our the fitting formulae may also become inaccurate. However, this limitation is mitigated by the fact that the 
cooling radius is larger at lower ambient densities. For example, for $\bar{n}_{\rm H}=10^{-2}$ cm$^{-3}$, $R_{\rm c}\sim 300$ pc, so that the simulation code will 
self-consistently compute the evolution of SNRs if the energy injection radius is less than this cooling radius by some margin. In Lagrangian codes with a 
mass resolution limit, the simulation must also satisfy the mass resolution requirement $\sim 300 M_{\odot}$ to capture the mass swept up during the energy-conserving phase of the SNR.}

\section{Testing the sub-resolution model with multiple SN simulations}
\label{sec:testing_subgrid}

In this section, we test our sub-resolution model for SN feedback using simulations of multiple SNe interacting self-consistently in a periodic box. We compare two simulations which differ primarily in the radius within which 
energy and momentum are injected (\S~\ref{sec:multiple_SNe_descr}, Table~\ref{tab:multiple}). Both simulations have a box size of 50 pc, a mean 
density $\bar{n}_{\rm H}=100$ cm$^{-3}$, solar metallicity, and a SN rate 
$\dot{n}_{\rm SN}=2\times 10^{-4}$ SNe Myr$^{-1}$ pc$^{-3}$, which corresponds to 25 SNe Myr$^{-1}$ in the box
(see \S \ref{sec:multiple_SNe_descr} for more details).
For our `sub-resolution' simulation, we inject radial momentum and thermal energy in a sphere of radius 8 pc following the fitting formulae of \S \ref{sec:fits}.  We use the formulae appropriate for a homogeneous medium 
because there is not much substructure within the injection region (this primarily reflects the limited physics in these simulations, rather than the true substructure of the ISM on scales of $\sim 8$ pc).  With an injection 
radius of 8 pc, we only resolve the cooling radius of 2 SNe out of the 50 exploding in the box, i.e. for 48 SNe the sub-grid prescription is used. For comparison, in the `resolved' simulation, the resolution (a factor 2 
better along each dimension) and the smaller injection region (radius 0.59 pc) allow us to resolve the cooling radius of all the SNRs with at least 14 cells.

\begin{figure*}
\includegraphics[width=0.99\textwidth]{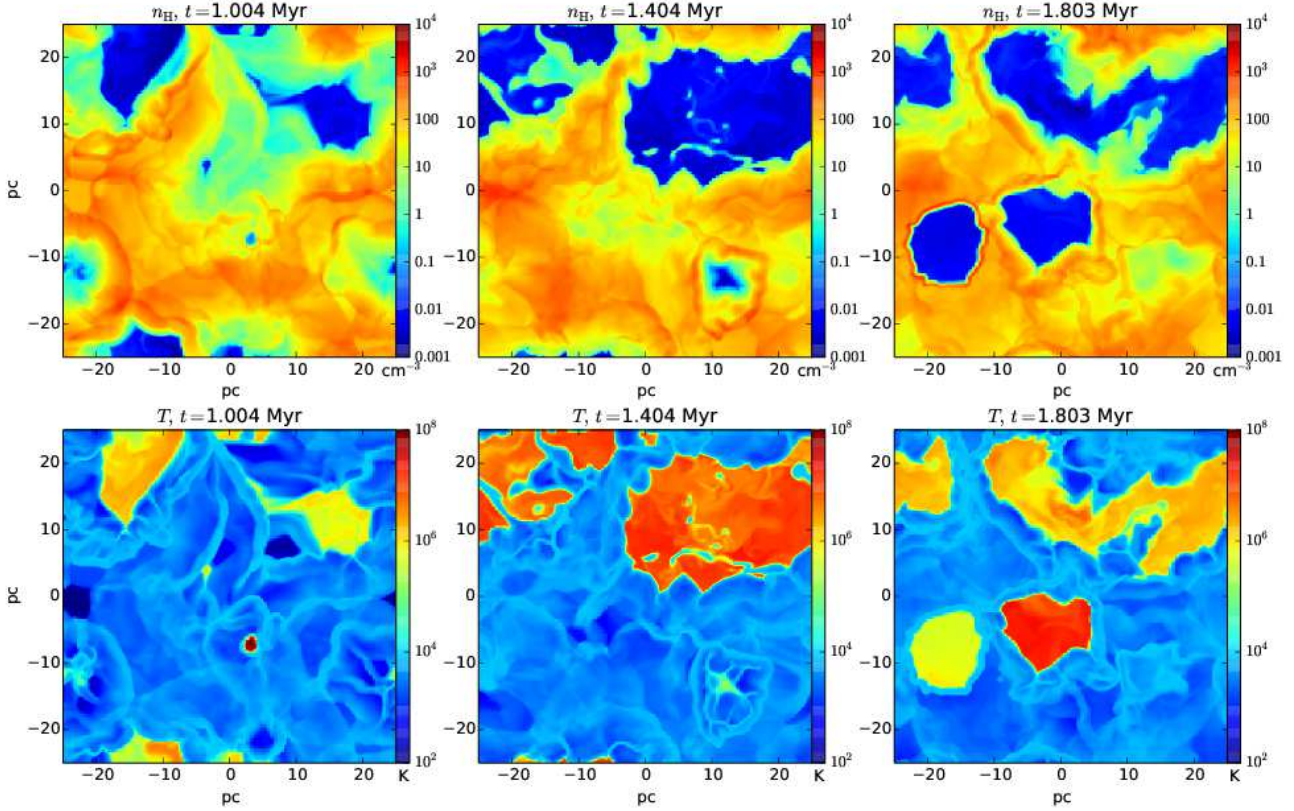}
\caption{Density (top) and temperature (bottom) maps for the `resolved' multiple SN simulation at different phases of its evolution.  In this simulation, energy is injected at sufficiently small radii that 
the key evolutionary phases of SNRs are self-consistently resolved.     The maps show a slice of thickness 5 pc through the centre of the 50 pc box.  The mean density of the box is $\bar{n}_{\rm H}=100$ cm$^{-3}$, the 
metallicity is $Z=Z_{\odot}$ and the supernova rate is $2\times 10^{-4}$ SNe Myr$^{-1}$ pc$^{-3}$, i.e., 25 SNe Myr$^{-1}$ in the box.}\label{mulsnr_map}
\end{figure*}

Figure~\ref{mulsnr_map} shows gas density and temperature maps in 5 pc thick slices for three snapshots of the ``resolved'' simulation.  The phase structure of the gas is complex and shows the effects of many shocks 
interacting with each other.  There are large cavities filled with hot, low density gas which are the result of SNe explosions.  These cavities survive for more than 0.1 Myr only if several SNRs overlap with each other or 
if a new SN explodes in an already under-dense region.

\begin{figure*}
  \includegraphics[width=0.95\textwidth]{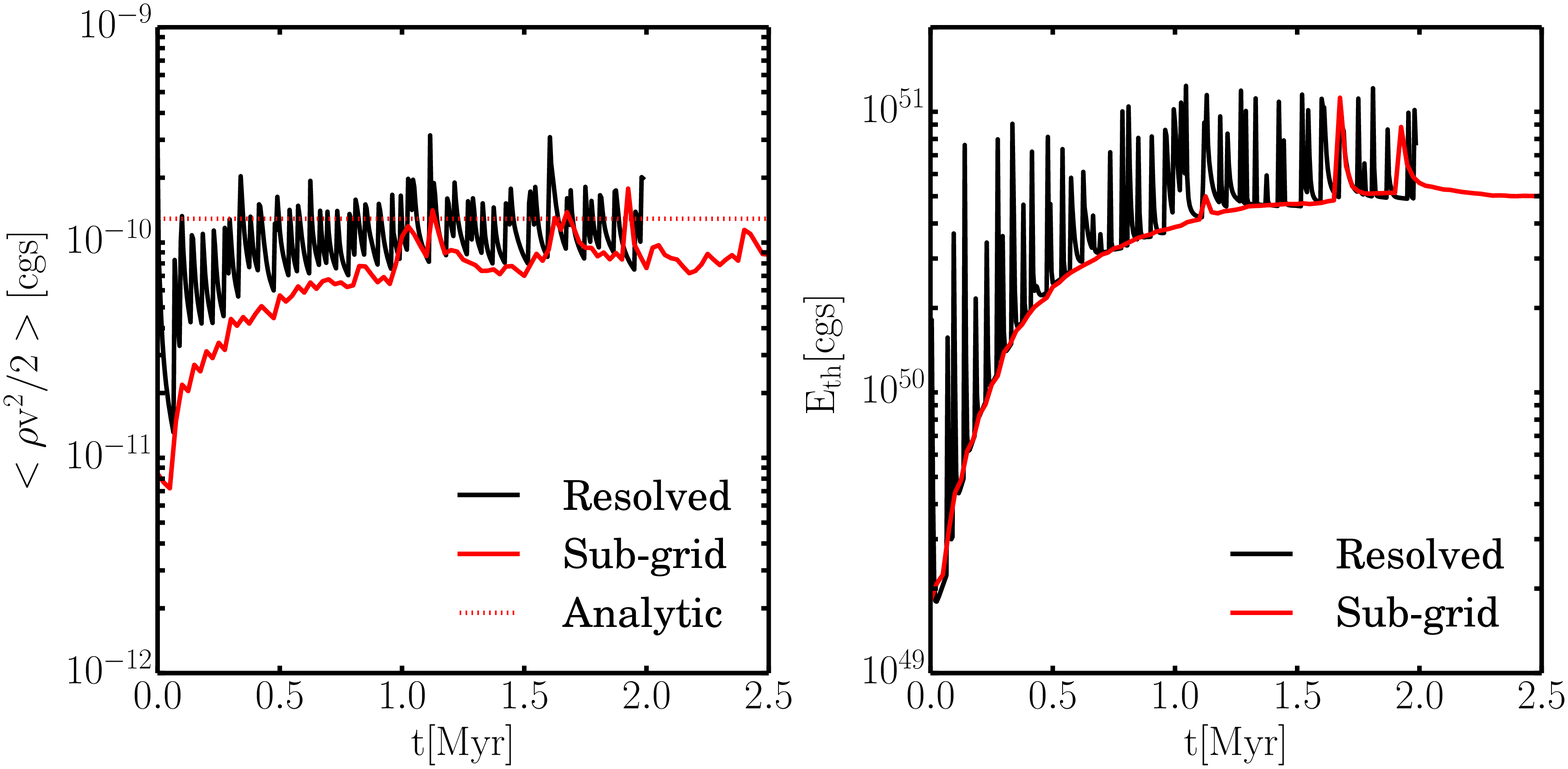} \caption{Evolution of the mass averaged turbulent kinetic energy density ({\em left panel}) and total thermal energy ({\em right panel}) in the multiple SN
  simulations. The solid black curves show the ``resolved'' simulation in which the Sedov-Taylor phase of the SNR is self-consistently resolved, while the solid red curves show the lower-resolution simulation in which SNe 
  energy and momentum are injected over a larger volume using the sub-grid 
  model for SN feedback proposed in \S \ref{sec:subgrid}.  The dotted line in the left panel shows an analytic approximation for the turbulent kinetic energy density (equation (\ref{sigma}) with $f=0.6$), described in 
  \S \ref{sec:testing_subgrid}. 
  Our sub-grid model accurately captures the dynamics of multiple interacting SNe in the fully resolved simulation. }\label{mulsnr}
\end{figure*}

Figure~\ref{mulsnr} shows the time evolution of the mass-averaged turbulent kinetic energy density (left panel) and thermal energy (right panel) in the periodic boxes with multiple SNe. 
The ``resolved'' simulation reaches a steady state after $\sim 1$ Myr. 
The sub-resolution model reproduces well the overall evolution of the thermal energy in the box. 
The turbulent kinetic energy density is slightly under-predicted by the sub-resolution model, but only by $\sim50\%$ after a steady state is reached. 
The evolution of the thermal and kinetic energy is smoother in time in the sub-resolution model than in the ``resolved'' simulation because the former generally does not capture the early time dynamics of individual SNRs. 
The exception is when a SN explodes in a low density region such that the cooling radius is larger than 8 pc.  In that case, even the sub-resolution model explicitly captures the evolution of the SNR.  This explains, 
in particular, the few spikes in thermal energy exhibited by the sub-resolution model in Figure \ref{mulsnr} (e.g., at $t\sim1.7$ Myr).  In addition to the mass averaged quantities compared in Figure \ref{mulsnr}, we also 
compared the resolved and sub-resolution model for the volume averaged turbulent velocity and thermal energy.  The agreement is worse there because the comparison is more sensitive to the early high velocity and temperature 
phases of SNRs that are (intentionally) not captured by the sub-grid simulation.

To test the robustness of our sub-grid model we also ran two additional sub-grid simulations in which the radii of the injection regions are 6 pc and 12 pc, respectively (instead of the 8 pc radius used in 
Figure \ref{mulsnr}). The results for an injection radius of 6 pc are consistent with the standard simulation. For the injection radius of 12 pc, however, we find that the turbulent kinetic energy density and thermal 
energy are underestimated by $\sim 50 \%$. This is because if the injection region is too large, the energy/momentum injected per cell will be comparable to the energy/momentum already present in the cell and the sub-grid 
prescription will not produce the strong shocks that should be associated with SNRs.  This indeed starts to be the case for an injection radius of 12 pc given the parameters of our multiple SN simulations.  More generally, 
the size of the injection region needs to be chosen so that the energy/momentum density injected per cell is greater than the local energy/momentum density per cell.

The turbulent kinetic energy in our simulations is also reasonably consistent with previous analytic estimates (e.g., \citealt{2011ApJ...731...41O,2013MNRAS.433.1970F}), which we summarize and refine somewhat here.  
We assume that SNe drive turbulence on a scale $L_{\rm drive}<L_{\rm box}$ and that SNRs expand until their velocity becomes comparable to the turbulent velocity dispersion in the ISM $\sigma$. 
We denote the final radius of SNRs when they merge with the turbulent ISM by $R_{\rm SN}$ and their velocity at that time by $V_{\rm SN}$. We assume that $L_{\rm drive} \sim R_{\rm SN}$ and $\sigma \sim V_{\rm SN}$. 
{The former is not guaranteed and depends in part on how efficiently SNR overlap with each other ($L_{\rm drive}\lesssim 100$ pc in general \cite{2006ApJ...653.1266J})}. We neglect overlap in our estimates.
Given this assumption, we can compute  $V_{\rm SN}$ from the asymptotic momentum of the SNR (assuming that the asymptotic momentum is reached before the SNR merges with the ISM):
\begin{equation}
V_{\rm SN}=\frac{3P_{\rm fin}}{4\pi \rho R_{\rm SN}^3}
\end{equation}
so that 
\begin{equation}\label{etwo}
R_{\rm SN}=\left(\frac{3P_{\rm fin}}{4\pi \rho V_{\rm SN}}\right)^{1/3}.
\end{equation}
In steady state the energy injection associated with SNe having momentum $P_{\rm fin}$ and velocity $V_{\rm SN} \sim \sigma$ balances the turbulent energy decay rate 
\begin{equation}
\frac{1}{2}\rho \sigma^2 \frac{\sigma}{L_{\rm drive}} \sim f P_{\rm fin} \dot{n}_{\rm SN} \sigma,
\label{edot}
\end{equation}
where the factor $f$ accounts for momentum cancellation when multiple blast waves interact and $\dot{n}_{\rm SN}$ is the SN rate per unit volume. 
Again setting $L_{\rm drive} \sim R_{\rm SN}$ and $\sigma \sim V_{\rm SN}$, equations (\ref{etwo}) and (\ref{edot}) imply
\begin{equation}
\sigma\approx \frac{3}{4\pi} \left( \frac{32 \pi^2}{9} \right)^{3/7}\left( \frac{P_{\rm fin}}{\rho}\right)^{4/7} \left( f \, \dot{n}_{\rm SN} \right)^{3/7}
\label{sigma}
\end{equation}
which allows to easily estimate the turbulent kinetic energy density $\rho \sigma^2/2$. 

For the multiple SN simulation shown in Figures \ref{mulsnr_map} \& \ref{mulsnr}, we find consistency between the analytic estimate in equation (\ref{sigma}) and the numerical results for $f \sim 0.6$ (using $P_{\rm fin}$ 
from the isolated SNR simulations for $\bar{n}_{\rm H} = 100$ cm$^{-3}$).  This implies that order unity of the momentum injected by SNe contributes to exciting turbulent motions. Note also that $R_{\rm SN}\sim 10$ pc for 
these simulations so that the turbulent crossing time of $R_{\rm SN}$ is $\sim 1$ Myr, similar to the time over which the simulations reach steady state.  

\subsection{Comparison to Other Sub-Resolution Models}
\label{sec:other-subgrid}

A common approach to modeling SN feedback is to suppress gas cooling for a specified period of time after $\sim 10^{51}$ erg of kinetic or thermal energy is injected into the 
ISM \citep[e.g.,][]{2006MNRAS.373.1074S, 2007MNRAS.374.1479G, 2010MNRAS.407.1581S, Governato2010,2011ApJ...742...76G}.  This is intended to mimic the finite time it takes radiative losses to set in during the evolution 
of SNRs.  \citet{2006MNRAS.373.1074S} proposed two classes of models, one in which cooling is shut off for a fixed time $\sim 10-30$ Myr.  Such models would obviously not be capable of correctly reproducing the turbulent 
or thermal energy in our multiple SN simulations, which reach a steady state after only $\sim 1$ Myr.  An improvement on this model is to suppress cooling over a timescale that depends explicitly on the local ISM 
properties, as in the second set of models developed by \citet{2006MNRAS.373.1074S}.  They advocate depositing a fraction of the $10^{51}$ erg per SNe into the ambient ISM and then shutting off cooling for a 
timescale $t_{\rm max} = 10^{6.85} E_{51}^{0.32} n^{0.34} P_{4}^{-0.7}$ yr, where $E_{51}$ is the SN energy in units of $10^{51}$ erg, $n$ is the ambient density, and $P_4$ is the ambient ISM pressure in units 
of $10^4 k$ cm$^{-3}$ K (subsequent work has deposited nearly the full $10^{51}$ ergs per SN associated with this feedback; e.g., \citealt{2011ApJ...742...76G,2014ApJ...792...99S}).  Explicitly evaluating $t_{\rm max}$ in 
our resolved multiple SN simulation, we find that $t_{\rm max}$ is at least a factor of 2-3 longer than the total duration of our simulation for all of the SNRs.  Thus the model advocated by \citet{2006MNRAS.373.1074S} 
(and used in a number of related studies; e.g., \citealt{Governato2010,2011ApJ...742...76G,2014ApJ...792...99S,Feldmann2014}) would overpredict by a factor of $\sim 100$ or more the thermal energy in our multiple SN 
simulations.  The reason for this lies in a misuse of the timescale $t_{\rm max}$ (see also \citealt{2013ApJ...770...25A}, who noted this as well).  The timescale over which most of the thermal energy of a SNR is 
radiated away is $\sim 10 t_c$, (with $t_c$ given in equations (\ref{tc-h}) and (\ref{tc-i})), which is a factor of $\sim 30$ shorter than $t_{\rm max}$.  This is because $t_{\rm max}$ defined in \citet{1977ApJ...218..148M} 
is roughly the timescale for the SNR to reach pressure equilibrium with its surroundings.  This is thus the timescale over which the {\em late-time residual momentum and thermal energy} of a SNR mixes with the ISM 
(the residual thermal energy is primarily associated with the reverse shock, not the forward shock).  The thermal energy content of this gas per SN is shown by the late time solutions in our 
Figures \ref{hom_vs_inh}-\ref{metdep} and is typically $\sim 0.3-3 \times 10^{49}$ erg.  Thus it is inconsistent to both deposit $\sim 10^{51}$ ergs per SN and shut off cooling for $\sim t_{\rm max}$.  
Allowing the Sedov-Taylor phase to effectively extend to $t_{\rm max}$ overestimates both the thermal energy and momentum injected into the ISM, the latter because $P_r \propto t^{3/5}$ in the Sedov-Taylor phase 
(given that $t_{\rm max}$ is a factor of $\sim 30$ larger than the true cooling time, the asymptotic momentum of a typical SNR is too large by a factor of $\sim 8$ in this model, at least for a roughly homogeneous medium).

\subsection{Comparison to Other Multiple Supernova Simulations}

\citet{2009ApJ...704..137J} carried out a set of three-dimensional simulations of multiple SN explosions in stratified galactic disks in which the gas surface density and SN rate are varied according to the Kennicutt-Schmidt 
law. In these simulations $10^{51}$ erg of thermal energy was injected in a sphere whose radius is typically $\sim 10-50$ pc depending on the local density; the injection radius is chosen such that radiative losses within the 
sphere are negligible for a few time steps \citep{2006ApJ...653.1266J}.  This does not, of course, guarantee that the proper SNR evolution is captured since this will depend on the size of the time step relative to the local 
cooling time of the SNR.  \citet{2009ApJ...704..137J} estimate a mass weighted 1D velocity dispersion $\sim 4-6$ km s$^{-1}$, relatively independent of SN rate -- this is comparable to the 1D turbulent velocity we find in our 
simulation.

\cite{2011ApJ...731...41O} and \cite{2012ApJ...754....2S} used analytic arguments and numerical hydrodynamical simulations of two-dimensional galactic disks to show that self-regulated star formation in high surface density 
disks can be achieved via the momentum input from SNe.  The authors carried out two dimensional simulations including the effects of gravity, Coriolis forces and stratification of the ISM that we neglect.  They used an 
isothermal equation of state and input solely momentum from SNe with $P_{\rm fin}/m_{*}\approx 3000$ km/s \citep{1998ApJ...500..342B}, which is somewhat larger than what we find in our isolated SNR simulations.  We have 
shown that high resolution simulations that capture more of the relevant physics (in particular, by resolving the Sedov-Taylor phase and the radiative cooling of SNRs) produce similar results to \cite{2012ApJ...754....2S}'s 
simulations with pure momentum injection.  In particular, both our and their simulations satisfy the basic energy balance given in our equation (\ref{edot}) with only modest `cancellation' of momentum between SNRs.  A more 
quantitative comparison is difficult at this stage because we did not explore a full range of SN rates, mean densities etc.

\cite{2014arXiv1411.0009G} used multiple SNe simulations in periodic boxes to study the properties of the self-consistently generated ISM. They considered different densities, SN rates and compared the properties of the ISM 
obtained when SN explode at density peaks (peak driving) or at random locations (random driving). They concluded that hot, low density phase of the ISM can form only if $\sim 50$ \% of the SN explode at random location in the 
box, i.e. preferentially in lower density regions.

In future work, we plan to extend our analysis of multiple SNe with higher-resolution simulations, larger boxes, different SN rates, and in stratified media where the competition between driving turbulence and 
driving galactic winds can be explicitly studied.   It will also be important to consider the spatial clustering of SNe explosions, since \cite{2014arXiv1405.7819H} showed that spatial clustering dramatically increases the 
effect of SN feedback on simulated Milky Way-like discs.


\section{Discussion and Conclusions}
\label{sec:summary}

We have used three-dimensional simulations of isolated supernova remnants (SNRs) to study their evolution as a function of ambient medium density and metallicity.  We have focused in particular on the radial momentum and 
thermal energy of the SNRs, since these quantities are particularly important for determining how SN feedback drives turbulence in the ISM and powers galactic winds, respectively. For a homogeneous ambient medium, our 
three-dimensional solutions agree reasonably well with previously published results using spherically-symmetric simulations \citep{1988ApJ...334..252C, 1998ApJ...500...95T}.  We extend these results to study the evolution 
of SNRs in an inhomogeneous medium with a lognormal density PDF -- statistics motivated by those of supersonic turbulence. For our inhomogeneous medium calculations, we varied the effective Mach number of the ambient medium 
and the maximum spatial scale of density fluctuations in addition to ambient density and metallicity.  Our main findings are as follows:

\begin{itemize} \item For low Mach numbers $\mathcal{M}\sim1$ characterizing the ambient medium density fluctuations, the latter are modest and SNRs evolve as in a homogeneous medium. In the limit $\mathcal{M}\gtrsim 10$, 
the evolution of SNRs is significantly affected by inhomogeneities in the ambient medium but the asymptotic radial momentum and thermal energy are not sensitive to the Mach number. The evolution of the thermal energy as a 
function of time is, however, sensitive to the Mach number and the characteristic spatial scale of the density fluctuations (Figure \ref{machdep}).  
\item At the same mean density, SNRs in an inhomogeneous medium propagate faster than in a homogeneous medium owing to the SN shock and hot shocked gas escaping through low density channels 
(Figures \ref{fig:maps}-\ref{fig:radius}).  As a consequence, the cooling radius in our simulations with an inhomogeneous ambient medium is a factor $\sim2$ larger than in the corresponding simulations with a homogeneous 
ambient medium. This implies much more efficient overlap of SNRs in an inhomogeneous medium since the volume per SNR prior to significant radiative losses is a factor of $\sim 10$ larger.  

\item The asymptotic radial momentum reached by SNRs is lower in an inhomogeneous medium by $\sim 30\%$ than in a homogeneous medium of the same mean density due to increased radiative losses from mixing of cool clouds 
with hot shocked gas (Figure \ref{rhodep}, Table \ref{tab:momentum}).  

\item If the spatial size of the largest density fluctuations in the ISM is comparable to the cooling radius of the SNR, the early evolution of the SNR is subject to stochastic effects related to whether the SN explodes 
in a local underdensity or overdensity (Figures \ref{fig:mapslc} and \ref{clumpdep}). If, however, the spatial size of the largest density fluctuations in the ISM is small compared to the cooling radius, density fluctuations 
average out.  In that case, the asymptotic radial momentum and thermal energy of the remnant are nearly independent of the maximum clump size.  The natural way to interpret these results is that SNR evolution is sensitive 
primarily to the mean ISM density within the cooling radius and not to the mean ISM density averaged over larger scales (since the remnant reaches those larger scales only when it is already momentum conserving).

\item The residual thermal energy at the end of SNR expansion decreases as a function of the mean density of the ISM (Figure \ref{rhodep}). The residual thermal energy is typically lower than the kinetic energy by a factor 
of a few (e.g., Figure \ref{hom_vs_inh}). In the limit that SNe do not overlap significantly and break out of the galactic disc, SN feedback is thus primarily `momentum-driven' rather than thermally driven. The effect of 
overlap and breakout is not captured by our individual SNR calculations.  
\end{itemize}

We have used our simulations of isolated SNRs to develop fitting formulae that capture the dependence of SNR evolution on ISM properties (\S \ref{sec:fits} ).  We then showed how these fitting formulae can be used to model 
SN feedback in simulations that lack the resolution to capture the key dynamical phases of SNRs (\S \ref{subgrid}).  Specifically, our proposed sub-grid model is one in which the ratio of thermal to kinetic energy injected 
into the ambient medium per SN depends on whether the cooling radius of the SNR is resolved.  Our proposed model is analogous to, but somewhat more accurate than, the sub-resolution model for SNe used in the FIRE (Feedback 
In Realistic Environments) cosmological simulation project \citep[][]{2014MNRAS.445..581H}. Given current and foreseeable resolution, this sub-grid model should be applicable to most cosmological simulations, including high 
resolution zoom-in simulations, and even to some non-cosmological simulations of isolated galaxies.

As a test of our proposed sub-grid model, we carried out simulations of multiple SNe interacting self-consistently in a periodic box of size (50 pc)$^3$ and mean density $\bar{n}_{\rm H}=100$ cm$^{-3}$. We explicitly 
compared the results of simulations in which the full $10^{51}$ erg of energy was injected on scales much smaller than the cooling radius (the ``right'' answer) to simulations that used the sub-grid model and injected SN
feedback on scales comparable to or larger than the typical cooling radius of the SNRs. The main results of our multiple SN simulations are:
\begin{itemize}
\item The statistical properties of the ISM reach an approximate steady state after $\sim 1$ Myr for both the resolved simulation and for SNRs approximated with our sub-resolution model (Figure \ref{mulsnr}). This time scale 
is comparable to the turbulent crossing time of the `outer scale' set by SNe.  \item The sub-resolution model accurately predicts the mass-averaged turbulent and thermal energy densities measured in the resolved simulation 
(Figure \ref{mulsnr}).
\item A fraction of order unity of the momentum injected by SNe in the medium is converted to turbulent motions. Simple analytic estimates of the rms turbulent velocity generated by SNe thus do a good job reproducing the 
simulation results (equation (\ref{sigma})).  

\item We argue that many of the `delayed cooling' schemes for SN feedback proposed and utilized in the literature \citep[e.g.,][]{2006MNRAS.373.1074S, Governato2010, 2014ApJ...792...99S} use an incorrect timescale for 
suppressing radiative cooling in SNRs (\S \ref{sec:other-subgrid}).  Such sub-grid models significantly overpredict the late-time thermal energy and momentum of SNRs.     
\end{itemize}

More generally, we advocate that stellar and black hole feedback models utilized in galaxy-scale or cosmological simulations should (as much as possible) be tested and refined using higher resolution smaller-volume 
simulations that capture more of the key physics (as in the present paper and, e.g., work on radiation pressure feedback by \citealt{KT2012,Davis2014}).  This approach is both computationally feasible and necessary to 
develop more predictive galaxy-scale and cosmological simulations.

There are many opportunities to extend the present study.  These include modeling additional physics such as thermal conduction, magnetic fields, and cosmic rays.  It would also be interesting to study how SN feedback 
interacts with other feedback mechanisms.  Galaxy-scale \citep[e.g.,][]{2012MNRAS.421.3488H, 2013ApJ...770...25A} and cosmological simulations \citep[e.g.,][]{2014MNRAS.445..581H, 2014arXiv1404.2613A} that implement 
approximations for multiple stellar feedback mechanisms show that they can interact non-linearly and in a non-trivial fashion.  In particular, it would be useful to understand how pre-processing of the ambient medium 
by stellar radiation and stellar winds affects SN feedback.  Spatial and temporal correlations of SNe may also significantly affect their consequences \citep[e.g.,][]{2014arXiv1405.7819H}. 
{In future work, we plan to study the effects of such correlations in the more realistic setting of vertically stratified galactic discs, which will allow us to better understand the formation of superbubbles and 
the generation of galactic winds. }

{Following the submission of our original manuscript, several other studies on SN feedback have become available. \cite{2014arXiv1410.1537K} also 
performed hydrodynamic simulations of SNe in an inhomogeneous medium. Their results are broadly consistent with ours where they overlap despite their use 
of an ambient medium shaped by the thermal instability instead of the lognormal model used in this paper. This demonstrates that the momentum injected by SNe is relative insensitive to the details of the ambient density 
structure of the ISM. \cite{2014arXiv1410.0011W} used smooth particle hydrodynamics simulations to 
study the effects of SNe on molecular clouds and reported momentum boost values comparable to those found in this paper (considering the effects of 
pre-processing of the medium by ionizing radiation, they found boosts enhanced by a factor $\sim 50$ \%). Finally, \cite{2014arXiv1410.3822S} and 
\cite{2015arXiv150105655K} independently implemented sub-grid SN models similar to the one proposed in this paper and introduced in \cite{2014MNRAS.445..581H}, 
though they did not investigate the effects of inhomogeneities in the ambient medium.}

\section*{Acknowledgments}
We thank Evan Scannapieco and Robert Feldmann for many useful discussions.  Davide Martizzi is a post-doctoral fellow at University of California, Berkeley and is funded by the Swiss National Science Foundation.  
Claude-Andr\'{e} Faucher-Gigu\`{e}re was supported by NASA through Einstein Postdoctoral Fellowship Award number PF3-140106, by NSF through grant number AST-1412836, and by Northwestern University funds.  This work was 
supported in part by NASA Grant ATP12-0183 and by a Simons Investigator Award to Eliot Quataert from the Simons Foundation. We thank our anonymous referee for his/her comments that greatly improved the quality of the paper.


\bibliography{paper_1.3}


\appendix
\section{Resolution test}\label{appendix:A}

\begin{figure*}
   \includegraphics[width=0.99\textwidth]{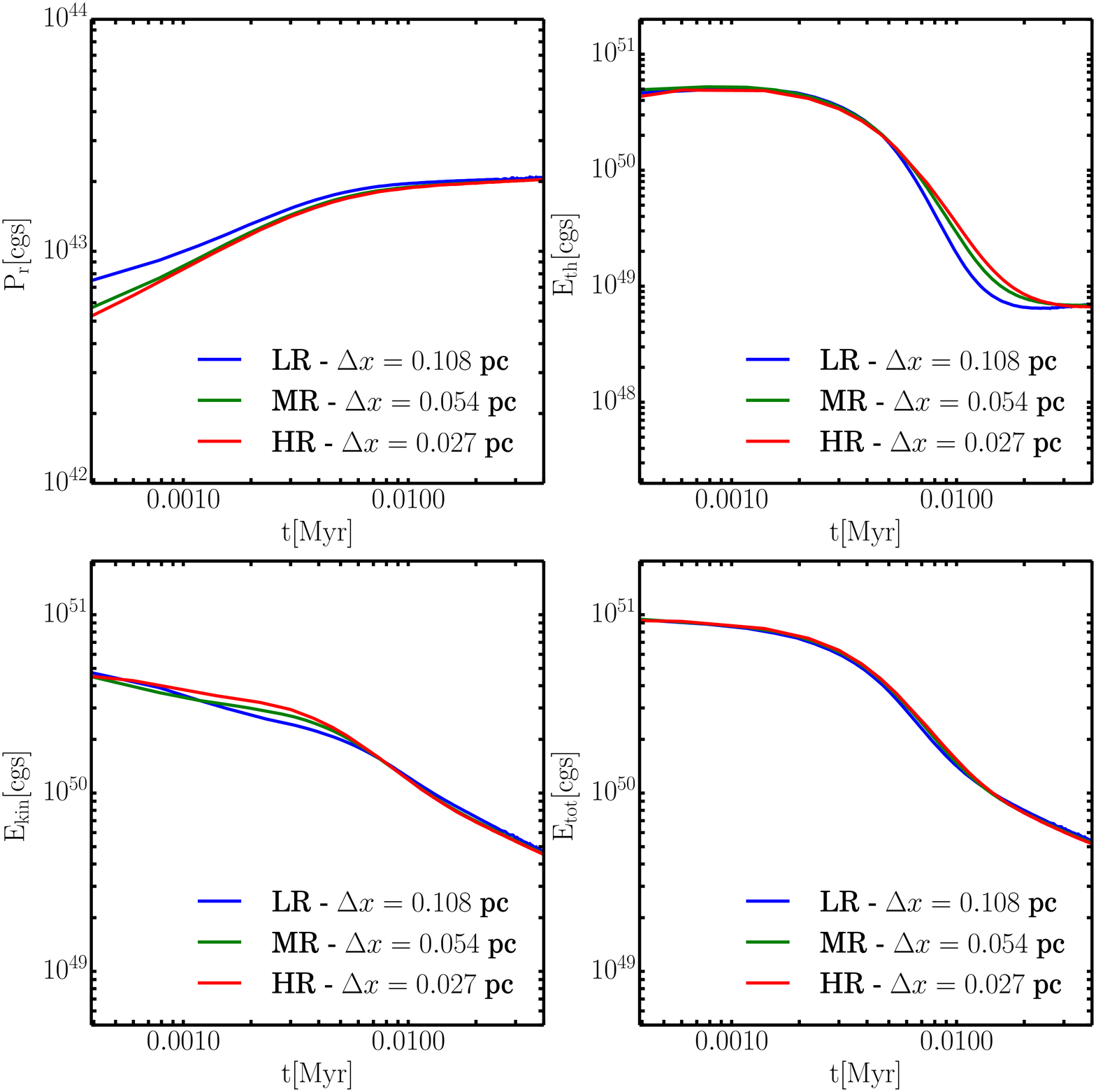}
\caption{ Resolution test for SNR evolution in an inhomogeneous medium of density $\bar{n}_{\rm H} = 100$ cm$^{-3}$, metallicity $Z = Z_{\odot}$ and Mach number 
$\mathcal{M}=30$. Top-left: radial momentum in the swept up gas. Top-right: thermal energy. Bottom-left: kinetic energy. Bottom-right: total energy. 
All quantities are well converged at the fiducial resolution of the simulations in this paper (${\rm MR}$, $512^3$ effective grid).}\label{fig:restest}
\end{figure*}

We have performed resolution tests to verify that our numerical solutions are converged.  One of these tests for a simulation of an isolated SNR in an inhomogeneous medium is shown in Figure~\ref{fig:restest}.  
The fiducial resolution we adopt throughout the paper is labeled ${\rm MR}$ ($512^3$ effective grid).  We also consider a simulation with resolution lower by a factor 2 in each dimension (${\rm LR}$; $256^{3}$ effective grid) and a simulation with higher resolution (${\rm HR}$; $1024^{3}$ effective grid).  
The plot shows that the evolution of momentum and energy (total, kinetic, and thermal) is well converged.


\label{lastpage}
\end{document}